\documentclass[12pt, titlepage]{article}
\usepackage[utf8]{inputenc}

\usepackage[comma,authoryear]{natbib}

\usepackage[english]{babel}
\usepackage{amsthm}
\theoremstyle{definition}

\theoremstyle{remark}

\usepackage{mathrsfs,amsmath}   
\usepackage{amsmath,amssymb}
\usepackage{bbm}
\usepackage{mathtools}

\usepackage{footnote}
\usepackage{footmisc}
\makesavenoteenv{tabular}
\makesavenoteenv{table}
\renewcommand{\thefootnote}{\alph{footnote}}

\newcommand{\astfootnote}[1]{
\let\oldthefootnote=\thefootnote
\setcounter{footnote}{0}
\renewcommand{\thefootnote}{\fnsymbol{footnote}}
\footnote{#1}
\let\thefootnote=\oldthefootnote
}

\usepackage{commath}
\usepackage{nicefrac}
\usepackage{titling,lipsum}
\usepackage{algpseudocode}
\usepackage{algorithm}
\usepackage{subfig}

\usepackage{hyperref}
\usepackage{xcolor}
\usepackage{sectsty}
\definecolor{MyBlue1}{rgb}{0,0,1}
\definecolor{MyBlue2}{rgb}{0.1,0.3,0.8}
\definecolor{MyBlue3}{rgb}{0.3,0.5,1}
\definecolor{MyGreen}{rgb}{0.4,1,0.4}
\definecolor{MyGreen2}{rgb}{0,0.6,0}
\definecolor{MyRed}{rgb}{1,0.4,0.4}
\hypersetup{
    colorlinks=true,
    linkcolor=MyBlue1,
    citecolor=MyBlue1,
    linkbordercolor=MyRed, 
    citebordercolor=MyGreen, 
    }

\usepackage{multirow}
\numberwithin{equation}{section}
\usepackage{cases}

\title{ \bf
Multivariate L\'evy models: calibration and pricing
}
\author{
{\bf Giovanni Amici}  \\
       Department of Mathematical Sciences\\
       Politecnico di Torino\\
       giovanni.amici@polito.it\\[14pt]
{\bf Paolo Brandimarte}  \\
       Department of Mathematical Sciences\\
       Politecnico di Torino\\[14pt]
{\bf Francesco Messeri}  \\
       Model Development and Integration Senior Specialist\\
       Intesa Sanpaolo Risk Management IMI CIB\\[14pt]
{\bf Patrizia Semeraro}  \\
       Department of Mathematical Sciences\\
       Politecnico di Torino\\
       }

\begin{document}

\maketitle

\begin{abstract}
The goal of this paper is to investigate how the marginal and dependence structures of a variety of multivariate L\'evy models affect calibration and pricing.
To this aim, we study the approaches of \cite{luciano2010multivariate} and \cite{ballotta2016multivariate} to construct multivariate processes.
We explore several calibration methods that can be used to fine-tune the models, and that deal with the observed trade-off between marginal and correlation fit. 
We carry out a thorough empirical analysis to evaluate the ability of the models to fit market data, price exotic derivatives, and embed a rich dependence structure.
By merging theoretical aspects with the results of the empirical test, we provide tools to make suitable decisions about the models and calibration techniques to employ in a real context.
\vfil


\noindent \textbf{Disclaimer}: The views and opinions expressed belong solely to the authors and do not necessarily reflect the views or positions of Intesa Sanpaolo SpA.

\end{abstract}

\section{Introduction}

Non-Gaussian L\'evy processes have demonstrated to be of massive interest in a number of areas in which stochastic factors need to be modeled.
Most notably, they combine a high mathematical tractability with a sophisticated dynamics that allows for jumps.
The laws of these processes are usually specified through explicit, or semi-explicit, characteristic functions that, for example, allow for relatively fast model calibrations (see, e.g., \citealp{cont2006retrieving}).
Also, efficient simulation schemes are available (see, e.g., \citealp{broadie2006exact}), which are useful, among other purposes, to estimate sensitivities of financial options (see \citealp{glasserman2010sensitivity}).
As such, L\'evy processes are suitable candidates for pricing problems whose solutions require numerical methods.
A well-known case is represented by the valuation of exotic derivatives, among which we highlight the optimal stopping problem of early-exercise contracts (see, e.g., \citealp{boyarchenko2002perpetual}, and \citealp{li2013optimal}).

The presence of jumps is of great importance, for example, in modeling extreme scenarios and hedging discontinuous payoff functions.
As these types of issues are hardly tackled by the traditional sensitivity-based hedging, some authors have resorted to flexible optimization frameworks in which jump processes can be easily cast.
Among these, we mention the stochastic programming formulation of the hedging problem developed by \cite{camciota2009pricing}, which use the optimal portfolio solution to deduce the fair price of the option to be hedged.
\cite{villaverde2004hedging} and
\cite{blomvall2022reducing} use similar techniques in the context of hedging and portfolio management, and they specifically represent some of the risk factors via L\'evy processes.
Dynamic programming formulations in presence of jumps are provided by \cite{bertsimas2001hedging} and \cite{godin2016minimizing}.
Further decisions problems under L\'evy dynamics can be found in the risk-related studies of \cite{fu2012equilibruim} and \cite{tang2023structural}, and in the context of production management (see \citealp{chiarolla2015optimal}).

In many of the aforementioned works, extending the problems to higher dimensions is theoretically possible and of great importance, given that real-world risk factors are typically correlated.
However, developing sound and tractable multivariate L\'evy processes is not an easy task (see \citealp{bianchi2022welcome} for a non-exhaustive list of these models).
As opposed to their univariate versions, the construction of these processes needs to provide flexibility of marginal and dependence structures and deal with possible trade-offs between them, especially in the calibration stage.
%
%

Our work fits the above-reported strand of literature by studying multivariate L\'evy processes for the valuation of exotic derivatives. 
In particular,
we explore the two model constructions introduced by \cite{luciano2010multivariate} and \cite{ballotta2016multivariate}, respectively.
The process of Luciano and Semeraro is a multivariate version of famous one-dimensional time-changed Brownian motions, such as the Variance Gamma \citep{madan1998variance} or Normal Inverse Gaussian \citep{barndorff1997processes}, and is designed to keep one-dimensional  marginal returns in known distribution classes.
The construction is done by linear combination of subordinators, a technique that allows to model nonlinear dependence.
On the other hand, the model proposed by Ballotta and Bonfiglioli is constructed through a parsimonious two-factor linear representation of the log-return dynamics.
As discussed by the proposers, building the multivariate log-return process via linear combinations provides a flexible characterization and an intuitive economic interpretation of the models, which are broken down into an idiosyncratic and a systematic part.

However, the model of Luciano and Semeraro and that of Ballotta and Bonfiglioli have core theoretical differences.
The former is built to preserve marginal processes in a given class.
The latter, in its general formulation, has marginal processes belonging to unspecified classes, although proper convolution conditions can be imposed to change this feature. 
Convolution conditions were initially imposed in the article where the model has been introduced, but then removed in subsequent works (see \citealp{ballotta2017multivariate} and \citealp{ballotta2019estimation}), as the authors asserted that such conditions were unnecessary to improve tractability.
Also, the construction proposed by Luciano and Semeraro allows to model nonlinear dependence separately from correlation.
As such, its margins can be mutually uncorrelated but still dependent.
On the contrary, when the margins of Ballotta and Bonfiglioli model are uncorrelated processes, they are also independent.

Besides the natural constructions of \cite{luciano2010multivariate} and \cite{ballotta2016multivariate} models, we examine alternative formulations that could provide some advantages in terms of calibration.
In particular, we explore a version of Luciano and Semeraro process that relaxes convolution constraints on subordinators.
In this way, model parameters lose some interpretability, but extend to wider domains, making model fitting potentially easier.
This approach, that lets the marginal log-return processes to be of unknown class, was taken for example by \cite{guillaume2013alpha}, which introduced an unconstrained version of the \textit{$\alpha$-Variance Gamma} model of \cite{semeraro2008multivariate}.
Also, in addition to considering the unconstrained Ballotta and Bonfiglioli model, in this work we study the original version that imposes margins to be of given class, noting that this construction can, for instance, reduce the computational effort of the calibration.
Overall, we analyze processes with constrained and unconstrained structures, and with Variance Gamma and Normal Inverse Gaussian underlying laws, for a total of eight models.

Our motivation for selecting the Gamma and the Inverse Gaussian as subordinators can be summarized in three points.
First, these processes belong to the celebrated class of tempered stable subordinators (see, e.g., \citealp{kuchler2013tempered}), which has been recognized to well-model the trading activity of financial assets.
Secondly, they are the only tempered stable subordinators with characteristic function in explicit form.
Moreover, a Brownian motion subordinated to a Gamma is a process with finite activity of jumps, as opposed to the Normal Inverse Gaussian which is of infinite activity.
Thus, by employing both designs we can assess whether or not a particular jump structure significantly outperforms the other, when it comes to fit equity log-returns.
%

As anticipated, a crucial issue to consider is model calibration.
In the context of option pricing, estimation needs to be carried out under the risk-neutral measure, which is why the model is typically fitted to quotes of liquidly-traded options. 
As a matter of fact, as the prices of liquid contracts can have different orders of magnitude (depending on their maturity and moneyness), it is a common practice to minimize the distances between the Black-Scholes volatilities implied from quoted prices and those implied from prices computed with the model of interest.
This procedure is, in fact, used to determine the marginal dynamics.
However, as multi-asset products are usually not liquid, it is hard to extract information about the expected joint dynamics of log-returns.
It is then customary to fit the dependence structure by matching the model correlation and a proxy of the market-implied correlation that is not necessarily recovered from prices of exchange-traded options.
%

Also, the procedure to fit the whole multivariate dynamics can follow different mechanisms.
It can be quite simple for constrained models, that allow for a two-step scheme: first, marginal parameters are fitted to liquid volatility surfaces, and then dependence parameters are found by matching model and market correlations.
This procedure has the practical advantage that marginal parameters can be estimated once and then applied to a range of different multi-asset products.
Also, handling multiple marginal calibrations is not heavy from a computational point of view.
However, marginal and dependence fits can be subject to a trade-off, especially in constrained models.
As such, we analyze some techniques that can be applied to distribute calibration errors on one side or the other.
Mainly, they consist of running \textit{joint} calibrations (i.e., calibrating correlation and marginals at the same time; see e.g., \citealp{marena2018pricing}), or allowing convolution conditions to be satisfied just up to a least-square approximation.

On the other hand, calibrating unconstrained models always requires the joint procedure, as dependence parameters are included in the marginal parametrization.
Joint calibration requires an optimization problem with a considerable number of variables, and can be then  more difficult to handle with respect to the two-step procedure.
However, as unconstrained models remove convolution restrictions on marginals, their joint calibration is relatively simplified by the exclusion of stricter constraints.
Most importantly, not only the more flexible parametrization of these models reduces the computational effort of joint calibration, it could also find solutions that would not be found in constrained models due to the restricted parameter domains.
To verify whether this happens, we assess the ability of unconstrained models to reduce the marginal vs correlation fitting trade-off in a real context.
Also, we discuss whether a more flexible parametrization justifies the suppression of convolution conditions from a global viewpoint. 

%
In designing the empirical test of the models, we consider a calibration setting tailored to our case study.
In particular, we extensively discuss our choices about the optimization algorithm, the market data, and the calibration procedures employed to fit processes.
As constrained models have more interpretable parametrizations and more homogeneous marginal structures among each other, we assess them primarily.
First, we calibrate such models at several trading dates, report marginal parameters, and provide a thorough comparison of the fitting performances.
Thereafter, we employ models to price selected \textit{worst-performance} derivatives issued by Intesa Sanpaolo bank and examine the ability of the models to replicate market-quoted prices.
We further perform a sensitivity analysis to show the effect of linear and nonlinear dependence on the valuation of exotics, rewarding models that better capture these features.
After that, we test unconstrained models and discuss their results alongside those of the constrained versions.
We finally highlight the implications of the empirical results at a more general level than the present application.
The paper is then outlined as follows.
In Section \ref{Model Theory}, we report the theory behind the multivariate models introduced by \cite{luciano2010multivariate} and \cite{ballotta2016multivariate} in their numerous versions.
In Section \ref{Calibration Methods}, we explore the risk-neutral calibration methods 
to fit multivariate processes.
In Section \ref{Worst Performance Derivatives}, we describe the exotic derivatives employed in the analysis.
We design the empirical test and discuss results in Section \ref{Empirical Analysis}, and draw conclusions in Section \ref{Conclusions}.


\section{Log-Return Processes}\label{Model Theory}

This section analyzes the approaches proposed by Luciano and Semeraro and by Ballotta and Bonfiglioli to construct multivariate processes.
We present such models in their multiple versions, preserving or not well-known marginal laws, and embedding Variance Gamma or Normal Inverse Gaussian components.
For each multivariate process, we report the time-1 marginal characteristic function and the correlation coefficient used for the calibrations described in Sections \ref{Calibration Methods} and \ref{Empirical Analysis}.
Appendix \ref{app:Models} reports a more detailed description of the models and points out some relevant features.
Also, we report the parametrizations in Table \ref{tab:model_parameters} and summarize some core features of the models in Table \ref{tab:model_features}, which provides an overview of the pros and cons of these constructions against each other and against comparable models in literature.
For sake of simplicity, for any stochastic process $\{A(t)\}_{t\geq0}$, we let $A := A(1)$ and recall that any L\'evy process has characteristic function fully derivable by their time-1 distribution.


\begin{table}[]
\scriptsize
\centering
\begin{tabular}{cc|c|c|c|}
\cline{2-5} 
& \multicolumn{4}{|c|}{LS}
\\
\cline{2-5} 
& \multicolumn{1}{|c|}{cVG} & uVG & cNIG & uNIG
\\
\hline
\multicolumn{1}{|l|}{Parameters}
& $\boldsymbol{\mu,\sigma,\kappa},a,\{\rho_{ij}\}_{i\neq j}$
& $\boldsymbol{\mu,\sigma,\kappa,\alpha},a,\{\rho_{ij}\}_{i\neq j}$
& $\boldsymbol{\beta,\delta,\gamma},a,\{\rho_{ij}\}_{i\neq j}$
& $\boldsymbol{\beta,\delta,\gamma,\alpha},a,\{\rho_{ij}\}_{i\neq j}$
\\
\hline
\multicolumn{1}{|l|}{\#parameters}
& $1+3n+n(n-1)/2$
& $1+4n+n(n-1)/2$
& $1+3n+n(n-1)/2$
& $1+4n+n(n-1)/2$
\\
\hline
\vspace{-0pt}
\\
\cline{2-5} 
& \multicolumn{4}{|c|}{BB}
\\
\cline{2-5} 
& \multicolumn{1}{|c|}{cVG} & uVG & cNIG & uNIG
\\
\hline
\multicolumn{1}{|l|}{Parameters}
& $\{\boldsymbol{\mu}_l, \boldsymbol{\sigma}_l, \boldsymbol{\kappa}_l\}_{l=\boldsymbol{X},Z}, \:\boldsymbol{b}$
& $\{\boldsymbol{\mu}_l, \boldsymbol{\sigma}_l, \boldsymbol{\kappa}_l\}_{l=\boldsymbol{X},Z}, \:\boldsymbol{b}$
& $\{\boldsymbol{\beta}_l, \boldsymbol{\delta}_l, \boldsymbol{\gamma}_l\}_{l=\boldsymbol{X},Z}, \:\boldsymbol{b}$
& $\{\boldsymbol{\beta}_l, \boldsymbol{\delta}_l, \boldsymbol{\gamma}_l\}_{l=\boldsymbol{X},Z}, \:\boldsymbol{b}$
\\
\hline
\multicolumn{1}{|l|}{\#parameters}
& $4n + 3$
& $4n + 3$
& $4n + 3$
& $4n + 3$
\\
\hline
\end{tabular}
\caption{
\small
Model parameters.
LS and BB stand for \cite{luciano2010multivariate} model and \cite{ballotta2016multivariate} model, respectively.
cVG (uVG) stands for constrained (unconstrained) Variance Gamma.
cNIG (uNIG) stands for constrained (unconstrained) Normal Inverse Gaussian.
}
\label{tab:model_parameters}
\end{table}

\begin{table}[]
\scriptsize
\centering
\begin{tabular}{cc|c|c|c|c|c|c|c|c|c|}
\cline{3-11} 
&& \multicolumn{4}{|c|}{LS} & \multicolumn{4}{c|}{BB}
&
Other TS
\\
\cline{3-10} 
&& cVG & uVG & cNIG & uNIG & cVG & uVG & cNIG & uNIG
&
subord.\
\\
\hline
\multicolumn{1}{|r|}{(I)}&
\multicolumn{1}{|l|}{Closed-form char.\ func.}
& \checkmark & \checkmark & \checkmark & \checkmark & \checkmark & \checkmark & \checkmark & \checkmark
& $\times$
\\
\hline
\multicolumn{1}{|r|}{(II)}&
\multicolumn{1}{|l|}{Finite/infinite variation}
& f.v. & f.v. & i.v. & i.v. & f.v. & f.v. & i.v. & i.v.
&
\\
\hline
\multicolumn{1}{|r|}{(III)}&
\multicolumn{1}{|l|}{$\rho_{\boldsymbol{Y}}=0 \:\not\Rightarrow \text{independence}$}
& \checkmark & \checkmark & \checkmark & \checkmark & $\times$ & $\times$ & $\times$ & $\times$
&
\\
\hline
\multicolumn{1}{|r|}{(IV)}&
\multicolumn{1}{|l|}{Margins of known class}
& \checkmark & $\times$ & \checkmark & $\times$ & $\approx$\checkmark & $\times$ & $\approx$\checkmark & $\times$
&
\\
\hline
\multicolumn{1}{|r|}{(V)}&
\multicolumn{1}{|l|}{Allows 2-step calibration}
& \checkmark & $\times$ & \checkmark & $\times$ & \checkmark & $\times$ & \checkmark & $\times$
&
\\
\hline
\multicolumn{1}{|r|}{(VI)}&
\multicolumn{1}{|l|}{Trade-off margins vs dep.}
& high & low & high & low & high & low & high & low
&
\\
\hline
\end{tabular}
\caption{
\small
Summary of the model features.
The table shows whether or not a model possesses the listed features (second column) through the symbols \checkmark and $\times$, respectively.
Rows (I)-(IV) compare the theoretical features of models of interest against each other (and against subordinated Brownian motions with generic tempered stable (TS) subordinators).
$\rho_{\boldsymbol{Y}}$ is the model correlation.
(V) and (VI) show calibration-related features.
LS and BB stand for \cite{luciano2010multivariate} model and \cite{ballotta2016multivariate} model, respectively.
cVG (uVG) stands for constrained (unconstrained) Variance Gamma.
cNIG (uNIG) stands for constrained (unconstrained) Normal Inverse Gaussian.
}
\label{tab:model_features}
\end{table}


\subsection{LS Models}\label{LS Models}

In this subsection we recap the main steps to build the factor-based processes introduced in \citep{luciano2010multivariate}, that we name \textit{LS processes}.
Let $\{\boldsymbol{B}(t)\}_{t\geq0}$ and $\{\boldsymbol{B}^{\rho}(t)\}_{t\geq0}$ be multivariate Brownian motions with L\'evy triplets $(\boldsymbol{\mu}, \boldsymbol{\Sigma}, \boldsymbol{0})$ and $(\boldsymbol{\mu}^{\rho}, \boldsymbol{\Sigma}^{\rho}, \boldsymbol{0})$, respectively. In particular,
$$
\boldsymbol{\mu}
=
\begin{pmatrix}
    \mu_1, ..., \mu_n
\end{pmatrix},
\hspace{20pt}
\boldsymbol{\Sigma}
=
\text{diag}
\begin{pmatrix}
    \sigma_1^2, ..., \sigma_n^2
\end{pmatrix},
\hspace{20pt}
\mu_j \in \mathbb{R},\: \sigma_j > 0,\: j=1,...,n.
$$
Consider further the multi-parameter $\boldsymbol{s}=(s_1,...,s_n)^T \in \mathbb{R}_+^n$ and the multi-parameter process introduced by \cite{barndorff2001multivariate}, corresponding to the above defined Brownian $\boldsymbol{B}(t)$, that is
\begin{equation}
\boldsymbol{B}(\boldsymbol{s})=\{(B_{1}(s_{1}),...,B_{n}(s_{n})),\boldsymbol{%
s}\in \mathbb{R}_{+}^{n}\},  \label{MultiBM}
\end{equation}%
with the partial order on $\mathbb{R}_+^n$,
\begin{equation*}
\boldsymbol{s}^{1}\preceq \boldsymbol{s}^{2}\,\,\,\Leftrightarrow
\,\,\,s_{j}^{1}\leq s_{j}^{2},\, \hspace{10pt} j=1,...n.
\end{equation*}%
Moreover, let $\{\boldsymbol{X}(t)\}_{t\geq0}$ be a multivariate subordinator, with independent components and $\{Z(t)\}_{t\geq0}$ be a subordinator, independent from $\boldsymbol{X}(t)$.

With these processes, we can then define the multivariate log-return process of Luciano and Semeraro as
\begin{equation}\label{Y}    
    \boldsymbol{Y}(t)
    =
    \boldsymbol{B} (\boldsymbol{X}(t)) +
    \boldsymbol{B}^{\rho} (Z(t))
    =
    \begin{pmatrix}
        B_1 (X_1(t)) + B^{\rho}_1 (Z(t))
        \\
        ...
        \\
        B_n (X_n(t)) + B^{\rho}_n (Z(t))
    \end{pmatrix},
\end{equation}
where $\boldsymbol{B}(\boldsymbol{X}(t))$ and $\boldsymbol{B}^{\rho}(Z(t))$ can be considered as the idiosyncratic and the systematic risk components of the dynamics of the assets. Also, we set
$$
\boldsymbol{\mu}^{\rho} =
\begin{pmatrix}
    \mu_1 \kappa_1, ..., \mu_n \kappa_n
\end{pmatrix}
$$
$$
\boldsymbol{\Sigma}^{\rho} =
\begin{pmatrix}
    \sigma_1^2 \kappa_1 & \rho_{12} \sigma_1 \sigma_2 \sqrt{\kappa_1} \sqrt{\kappa_2} & ... & \rho_{1n} \sigma_1 \sigma_n \sqrt{\kappa_1} \sqrt{\kappa_n}
    \\
    \rho_{12} \sigma_1 \sigma_2 \sqrt{\kappa_1} \sqrt{\kappa_2} & \sigma_2^2 \kappa_2 & ... & \rho_{2n} \sigma_2 \sigma_n \sqrt{\kappa_2} \sqrt{\kappa_n}
    \\
    ... & ... & ... & ...
    \\ 
    \rho_{1n} \sigma_1 \sigma_n \sqrt{\kappa_1} \sqrt{\kappa_n} & \rho_{2n} \sigma_2 \sigma_n \sqrt{\kappa_2} \sqrt{\kappa_n} & ... & \sigma_n^2 \kappa_n
\end{pmatrix},
$$
with $\kappa_j>0, j=1,...,n$,
in such a way that each marginal log-return process $j$ is a Brownian motion with drift $\mu_j$, diffusion $\sigma_j$, and subordinated
by a factor-based subordinator
\begin{equation}\label{G_j}
G_j(t) = X_j(t) + \kappa_j Z(t).
\end{equation}
In particular, \cite{luciano2010multivariate} proved that
\begin{equation}\label{Y_j}
Y_{j}(t) \overset{\text{d}}{=} \mu _{j}G_{j}(t)+\sigma _{j} \Tilde{B}_j(G_{j}(t)),
\end{equation}%
where $\Tilde{B}_j$ is a standard Brownian motion.

By means of Theorem 30.1 in \citep{ken1999levy}, and Theorem 3.3 in \citep{barndorff2001multivariate} for the multivariate case, we are able to derive the time-1 characteristic function of the subordinated process as
\begin{equation}\label{LS-CF}
\phi _{\boldsymbol{Y}}(\boldsymbol{u}) =
\exp \Bigg\{ \sum_{j=1}^{n}l_{X_{j}}(\psi _{B_{j}}(u_{j}))
+
l_{Z}(\psi _{\boldsymbol{B}^{\rho }}(\boldsymbol{u})) \Bigg\},
\hspace{10pt}
\boldsymbol{u} \in \mathbb{R}^n,
\end{equation}
where $l(\cdot)$ and $\psi(\cdot)$ are the Laplace exponent and the characteristic exponent, respectively.
Also, pairwise correlations are given by
$$
\rho_{\boldsymbol{Y}} (i,j)
=
\frac{
\mathbb{E} [B_i^{\rho}] \mathbb{E} [B_j^{\rho}] \mathbb{V} (Z)
+ \text{Cov} ( B_i^{\rho}, B_j^{\rho} ) \mathbb{E}[Z]
}{\sqrt{ \mathbb{V}(Y_i) \mathbb{V}(Y_j) }}
=
\frac{
\mu_i \mu_j \kappa_i \kappa_j \mathbb{V} (Z)
+ \rho_{ij} \sigma_i \sigma_j \sqrt{\kappa_i} \sqrt{\kappa_j} \mathbb{E}[Z]
}{
\sqrt{ \mathbb{V}(Y_i) \mathbb{V}(Y_j) }
},
$$
where $\mathbb{E}[\cdot]$, $\mathbb{V}(\cdot)$ and $\text{Cov}(\cdot, \cdot)$ denote the expectation, variance and covariance, respectively.

Given the general LS construction provided above, we now present the model specifications employed in the analysis.
We report their marginal characteristic functions and correlation coefficients, denoted with $\phi_{Y_j}(\cdot)$ and $\rho_{\boldsymbol{Y}}(i,j)$ respectively, in the list below.
The ranges and constraints of the parameters can be seen in Eqs.\ \eqref{Gamma-Sub}, \eqref{Gamma-Sub-unc}, \eqref{IG-Sub}, and \eqref{IG-Sub-unc}.
\begin{itemize}
    \item Constrained LS-Variance Gamma
    \begin{equation}\label{VG-CF}
    \phi_{Y_j} (u) = \left( 1 - i u \mu_j \kappa_j + \frac{1}{2} u^2 \sigma^2_j \kappa_j \right)
    ^{-\kappa^{-1}_j}
    \end{equation}
    \begin{equation}\label{rhoY-LS-VG}
    \rho_{\boldsymbol{Y}} (i,j)
    =
    \frac{ a \left(
    \mu_i \mu_j \kappa_i \kappa_j  +
    \rho_{ij} \sigma_i \sigma_j \sqrt{\kappa_i} \sqrt{\kappa_j}
    \right)
    }{
    \sqrt{ (\sigma^2_i + \mu^2_i \kappa_i)
    (\sigma^2_j + \mu^2_j \kappa_j) }
    }
    \end{equation}
    \begin{equation}\label{Gamma-Sub}        
    \mu_j\in\mathbb{R}, \hspace{5pt}
    \sigma_j>0, \hspace{5pt}
    \kappa_j>0, \hspace{5pt}
    0 < a < \min_j \left( 1 / \kappa_j \right), \hspace{5pt}
    \abs{\rho_{ij}}\leq1
    \end{equation}
    \item Unconstrained LS-Variance Gamma
    \begin{equation}\label{VG-CF-uncLS}
    \phi_{Y_j} (u) = \left( 1 - i u \mu_j \kappa_j + \frac{1}{2} u^2 \sigma^2_j \kappa_j \right)
    ^{-(\alpha_j+a)}
    \end{equation}
    \begin{equation}\label{rhoY-LS-VG-uncLS}
    \rho_{\boldsymbol{Y}} (i,j)
    =
    \frac{ a \left(
    \mu_i \mu_j \kappa_i \kappa_j +
    \rho_{ij} \sigma_i \sigma_j \sqrt{\kappa_i} \sqrt{\kappa_j}
    \right)
    }{
    \sqrt{ \kappa_i (\alpha_i + a) (\sigma^2_i + \mu^2_i \kappa_i) \:
    \kappa_j (\alpha_j + a)(\sigma^2_j + \mu^2_j \kappa_j) }
    }
    \end{equation}
    \begin{equation}\label{Gamma-Sub-unc}        
    \mu_j\in\mathbb{R}, \hspace{5pt}
    \sigma_j>0, \hspace{5pt}
    \kappa_j>0, \hspace{5pt}
    \alpha_j>0, \hspace{5pt}
    a>0, \hspace{5pt}
    \abs{\rho_{ij}}\leq1
    \end{equation}      
    \item Constrained LS-Normal Inverse Gaussian
    \begin{equation}\label{NIG-CF}
    \phi_{Y_j} (u) = \exp \left( -\delta_j \left(\sqrt{\gamma^2_j - (\beta_j + i u)^2} - \sqrt{\gamma^2_j - \beta^2_j}\right) \right)
    \end{equation}
    \begin{equation}\label{rhoY-LS-NIG}
    \rho_{\boldsymbol{Y}} (i,j)
    =
    \frac{ a \left(
    \beta_i \delta_i^2 \kappa_i
    \beta_j \delta_j^2 \kappa_j +
    \rho_{ij} \delta_i \sqrt{\kappa_i} \delta_j \sqrt{\kappa_j}
    \right)
    }{
    \sqrt{
    \gamma_i^2 \delta_i (\gamma_i^2 - \beta_i^2)^{-3/2}
    \:
    \gamma_j^2 \delta_j (\gamma_j^2 - \beta_j^2)^{-3/2}
    }
    }
    \end{equation}
    \begin{equation}\label{IG-Sub}        
    \delta_j>0, \hspace{5pt}
    \gamma_j>0, \hspace{5pt}
    \abs{\beta_j}<\gamma_j, \hspace{5pt}
    0 < a < \min_j \left( \delta_j \sqrt{\gamma_j^2 - \beta_j^2} \right), \hspace{5pt}
    \abs{\rho_{ij}}\leq1
    \end{equation}
    \item Unconstrained LS-Normal Inverse Gaussian
    \begin{equation}\label{NIG-CF-uncLS}
    \phi_{Y_j} (u) = \exp \left( -\delta_j \left(\sqrt{\gamma^2_j - (\beta_j + i u)^2} - \sqrt{\gamma^2_j - \beta^2_j}\right) \left( \alpha_j + a \sqrt{\kappa_j} \right) \right)
    ,
    \hspace{15pt}
    u \in \mathbb{R}
    \end{equation}
    \begin{equation}\label{rhoY-LS-NIG-uncLS}
    \rho_{\boldsymbol{Y}} (i,j)
    =
    \frac{ a \left(
    \beta_i \delta_i^2 \kappa_i
    \beta_j \delta_j^2 \kappa_j +
    \rho_{ij} \delta_i \sqrt{\kappa_i} \delta_j \sqrt{\kappa_j}
    \right)
    }{
    \sqrt{
    (\alpha_i + a \sqrt{\kappa_i})
    \gamma_i^2 \delta_i (\gamma_i^2 - \beta_i^2)^{-3/2}
    \:
    (\alpha_j + a \sqrt{\kappa_j})
    \gamma_j^2 \delta_j (\gamma_j^2 - \beta_j^2)^{-3/2}
    }
    }
    \end{equation}
    \begin{equation}\label{IG-Sub-unc}        
    \delta_j>0, \hspace{5pt}
    \gamma_j>0, \hspace{5pt}
    \abs{\beta_j}<\gamma_j, \hspace{5pt}
    \alpha_j>0, \hspace{5pt}
    a>0, \hspace{5pt}
    \abs{\rho_{ij}}\leq1
    \end{equation}
\end{itemize}
More detailed descriptions of the above specifications are provided by Appendices \ref{LS-VG} \ref{LS-VG-uncLS}, \ref{LS-NIG}, and \ref{LS-NIG-uncLS}.

\subsection{BB Models}

We are now ready to present the second class of models, introduced by \cite{ballotta2016multivariate}. The authors proposed a multivariate process constructed by a convolution of two L\'evy processes, without the need to pass through subordinators.
We call processes of this kind as \textit{BB processes}, that are of the form
$$
\boldsymbol{Y}(t)
=
\boldsymbol{X}(t) + \boldsymbol{b}
\hspace{2pt}
Z(t)
=
\begin{pmatrix}
    X_1(t) + b_1 Z(t)
    \\
    ...
    \\
    X_n(t) + b_n Z(t)
\end{pmatrix},
$$
where $\boldsymbol{b} \in \mathbb{R}^n$, and $X_1(t), ..., X_n(t), Z(t)$ are $\mathbb{R}$-valued mutually independent L\'evy processes. Time-1 characteristic function is then

\begin{equation}\label{BB-CF}
\phi_{\boldsymbol{Y}}(\boldsymbol{u})
=
\exp \left\{ \sum_{j=1}^{n} \psi_{X_{j}}(u_{j})
+
\psi_Z \left( \sum_{j=1}^{n} b_j u_j \right) \right\},
\hspace{10pt}
\boldsymbol{u} \in \mathbb{R}^n,
\end{equation}%
where $\psi(\cdot)$ denotes the characteristic exponent.
Also in this case, it is natural to consider $\boldsymbol{X}(t)$ and $Z(t)$ as the idiosyncratic and the systematic risk factors, respectively.
The pairwise correlations are given by
$$
\rho_{\boldsymbol{Y}} (i,j)
=
\frac{
b_i b_j \mathbb{V} (Z)
}{
\sqrt{ \mathbb{V}(Y_i) \mathbb{V}(Y_j) }
}
.
$$
As opposed to LS, BB models do not necessarily suffer trade-offs between marginal and dependence fit.
However, to obtain one-dimensional margins belonging to a known class (e.g., VG, NIG), we would have to first set the distributions for $Y_j(t), j=1,...,n$, and then imposing the following convolution conditions on marginal distributions,
\begin{equation}\label{conv-conditions}
\psi_{Y_j} (u) = \psi_{X_j} (u) + \psi_Z (b_j u)
,
\hspace{15pt}
j = 1, ..., n.
\end{equation}
Specifically, the fitting procedure of constrained BB models consists of first estimating $Y_j(t)$ parameters, and then finding combinations of $X_j(t)$, $b_j$ and $Z(t)$ parameters that reflect market correlation, while still being marginally distributed as $Y_j(t)$.
However, such convolution conditions can require restrictive constraints on parameters, likely to be satisfied only up to an approximation.

The studied specifications of the BB construction are indicated as follows.
Let $\phi_{Y_j}(\cdot)$ and $\rho_{\boldsymbol{Y}}(i,j)$ denote the marginal characteristic function and the correlation coefficient, respectively.
Then, the model specifications are reported as follows.
\begin{itemize}
    \item Constrained BB-Variance Gamma
    \begin{equation}\label{VG-CF-copyBB}
    \phi_{Y_j} (u) = \left( 1 - i u \mu_j \kappa_j + \frac{1}{2} u^2 \sigma^2_j \kappa_j \right)
    ^{-\kappa^{-1}_j}
    \end{equation}
    \begin{equation}\label{rhoY-BB-VG}
    \rho_{\boldsymbol{Y}} (i,j)
    =
    \frac{
    b_i b_j (\sigma^2_Z + \mu^2_Z \kappa_Z)
    }{
    \sqrt{ (\sigma^2_i + \mu^2_i \kappa_i)
    (\sigma^2_j + \mu^2_j \kappa_j) }
    }
    \end{equation}
\begin{align}
    \label{cBB-VG-ranges}
    &
    \hspace{-0.4cm}
    \mu_j, \mu_{X_j}, \mu_Z \in\mathbb{R}, \hspace{5pt}
    \sigma_j, \sigma_{X_j}, \sigma_Z >0, \hspace{5pt}
    \kappa_j, \kappa_{X_j}, \kappa_Z >0, \hspace{5pt}
    b_j>0
\\
    \label{VG-conv-restrictions-1}
    &
    \hspace{-0.4cm}
    \kappa_j \mu_j = \kappa_Z b_j \mu_Z, \hspace{15pt}
    \kappa_j \sigma^2_j = \kappa_Z b^2_j \sigma^2_Z
\\
    \label{VG-conv-restrictions-2}
    & 
    \hspace{-0.4cm}
    \mu_j = \mu_{X_j} + b_j \mu_Z,
    \hspace{15pt}
    \sigma_j = \sqrt{ \sigma^2_{X_j} + b^2_j \sigma^2_Z },
    \hspace{15pt}
    \kappa_j = \frac{ \kappa_{X_j} \kappa_Z }{ \kappa_{X_j} + \kappa_Z }
\end{align}
    %
    \item Unconstrained BB-Variance Gamma
    \begin{equation}\label{VG-CF-uncBB}
    \phi_{Y_j} (u) =
    \left( 1 - i u \mu_{X_j} \kappa_{X_j} + \frac{1}{2} u^2 \sigma^2_{X_j} \kappa_{X_j} \right)
    ^{-\kappa^{-1}_{X_j}}
    \hspace{1pt}
    \left( 1 - i u b_j \mu_Z \kappa_Z + \frac{1}{2} u^2 b_j^2 \sigma^2_Z \kappa_Z \right)
    ^{-\kappa^{-1}_Z}
    \end{equation}
    \begin{equation}\label{rhoY-BB-VG-uncBB}
    \rho_{\boldsymbol{Y}} (i,j)
    =
    \frac{
    b_i b_j (\sigma^2_Z + \mu^2_Z \kappa_Z)
    }{
    \sqrt{
    \left[ \sigma^2_{X_i} + \mu^2_{X_i} \kappa_{X_i} + b_i^2 (\sigma^2_Z + \mu^2_Z \kappa_Z ) \right]
    \left[ \sigma^2_{X_j} + \mu^2_{X_j} \kappa_{X_j} + b_j^2 (\sigma^2_Z + \mu^2_Z \kappa_Z ) \right]
    }
    }
    \end{equation}
\begin{equation}\label{uBB-VG-ranges}
    \mu_{X_j}, \mu_Z \in\mathbb{R}, \hspace{5pt}
    \sigma_{X_j}, \sigma_Z >0, \hspace{5pt}
    \kappa_{X_j}, \kappa_Z >0, \hspace{5pt}
    b_j>0
\end{equation}
    \item Constrained BB-Normal Inverse Gaussian
    \begin{equation}\label{NIG-CF-copyBB}
    \phi_{Y_j} (u) = \exp \left( -\delta_j \left(\sqrt{\gamma^2_j - (\beta_j + i u)^2} - \sqrt{\gamma^2_j - \beta^2_j}\right) \right)
    \end{equation}
    \begin{equation}\label{rhoY-BB-NIG}
    \rho_{\boldsymbol{Y}} (i,j)
    =
    \frac{
    b_i b_j \gamma^2_Z \delta_Z (\gamma^2_Z - \beta^2_Z)^{-3/2}
    }{
    \sqrt{ \gamma^2_i \delta_i (\gamma^2_i - \beta^2_i)^{-3/2}
    \:\:
    \gamma^2_j \delta_j (\gamma^2_j - \beta^2_j)^{-3/2} }
    }
    \end{equation}
\begin{align}
        \label{cBB-NIG-ranges}
    &
    \hspace{-0.4cm}
    \delta_j, \delta_{X_j}, \delta_Z >0, \hspace{5pt}
    \gamma_j, \gamma_{X_j}, \gamma_Z >0, \hspace{5pt}
    b_j>0, \hspace{5pt}
    \abs{\beta_A}<\gamma_A, \hspace{5pt}
    A = j, X_j, Z
\\
    \label{NIG-conv-restrictions-1}
    &
    \hspace{-0.4cm}
    \beta_j = b_j^{-1} \beta_Z,
    \hspace{5pt}
    \gamma_j = b_j^{-1} \gamma_Z
\\
    \label{NIG-conv-restrictions-2}
    &
    \hspace{-0.4cm}
    \beta_j = \beta_{X_j} = b_j^{-1} \beta_Z,
    \hspace{5pt}
    \delta_j = \delta_{X_j} + b_j \delta_Z,
    \hspace{5pt}
    \gamma_j = \gamma_{X_j} = b_j^{-1} \gamma_Z
\end{align}
%
    \item Unconstrained BB-Normal Inverse Gaussian
    \begin{equation}\label{NIG-CF-uncBB}
    \begin{split}
    \phi_{Y_j} (u) =
    & \exp \left( -\delta_{X_j} \left(\sqrt{\gamma^2_{X_j} - (\beta_{X_j} + i u)^2} - \sqrt{\gamma^2_{X_j} - \beta^2_{X_j}}\right) \right) \cdot
    \\
    \cdot
    & \exp \left( -\delta_Z \left(\sqrt{\gamma^2_Z - (\beta_Z + i u b_j)^2} - \sqrt{\gamma^2_Z - \beta^2_Z}\right) \right)
    \end{split}
    \end{equation}
    \begin{equation}\label{rhoY-BB-NIG-uncBB}
    \rho_{\boldsymbol{Y}} (i,j) =
    \frac{
    b_i b_j \gamma^2_Z \delta_Z (\gamma^2_Z - \beta^2_Z)^{-3/2}
    }{
    \underset{m=i,j}{\prod}
    \sqrt{\gamma^2_{X_m} \delta_{X_m} (\gamma^2_{X_m} - \beta^2_{X_m})^{-3/2}
    + b_m^2
    \gamma^2_Z \delta_Z (\gamma^2_Z - \beta^2_Z)^{-3/2}}
    }
    \end{equation}
\begin{equation}\label{uBB-NIG-ranges}
    \delta_{X_j}, \delta_Z >0, \hspace{5pt}
    \gamma_{X_j}, \gamma_Z >0, \hspace{5pt}
    \abs{\beta_{X_j}}<\gamma_{X_j}, \hspace{5pt}
    \abs{\beta_Z}<\gamma_Z, \hspace{5pt}    
    b_j>0
\end{equation}
\end{itemize}
where Eqs.\
\eqref{cBB-VG-ranges}, \eqref{VG-conv-restrictions-1}, \eqref{VG-conv-restrictions-2}, \eqref{uBB-VG-ranges},
\eqref{cBB-NIG-ranges}, \eqref{NIG-conv-restrictions-1}, \eqref{NIG-conv-restrictions-2}, and \eqref{uBB-NIG-ranges}
show the ranges and constraints of the parameters.
A more thorough overview of the above models is given by Appendices \ref{BB-VG} \ref{BB-VG-uncBB}, \ref{BB-NIG}, and \ref{BB-NIG-uncBB}.

\section{Calibration Methods}\label{Calibration Methods}

In this section, we illustrate the calibration methods that can be employed to fit multivariate models to market data.
In terms of asset prices, we calibrate an $\mathbb{R}^n$-valued asset price process $\{ \boldsymbol{S} (t) \}_{t \geq 0}$ with margins given by
$$
S_j (t) = \exp \{ (r - q_j + g_j) t + Y_j (t) \}
,
\hspace{15pt}
j = 1, ..., n
$$
where $Y_j(t)$ is the $j$-th marginal of any of the multivariate log-return processes $\{ \boldsymbol{Y} (t) \}_{t \geq 0}$ introduced in Section \ref{Model Theory}.
Also, $r$ is the risk-free rate, $q_j$ is the $j$-th continuously compounded dividend yield and $g_j = - \psi_{Y_j} (u = -i)$ is the $j$-th mean correction to assure the martingale condition, needed in the risk-neutral context.

As described by \cite{marena2018pricing} and \cite{ballotta2016multivariate}, when LS and BB models impose margins of known L\'evy class, they allow for a two-stage procedure: first, marginal parameters are calibrated by exploiting liquid volatility surfaces; then, dependence parameters are found by matching model and market correlations.

Marginal calibration consists of finding, for each asset $j=1,...,n$, the set $\mathcal{M}_j$ of marginal parameters as
\begin{equation}\label{marginal calibration}
\begin{aligned}
\mathcal{M}^*_j
=
\hspace{5pt}
& \underset{
\mathcal{M}_j
}{\text{argmin}}
& & \frac{1}{N_j} \sum_{l=1}^{N_{j}} \omega_l \big[ v_l^{\text{mod}} (\mathcal{M}_j) - v_l^{\text{mkt}} \big]^2
\\
& \text{subject to}
& & \mathcal{M}_j \in \overline{\mathcal{M}}_j
\end{aligned}
\end{equation}
where $v_l^{\text{mod}} (\mathcal{M}_j)$ and $v_l^{\text{mkt}}$, $l=1,\ldots,N_j$, are Black-Scholes implied volatilities from model and market prices, respectively, $\overline{\mathcal{M}}_j = \{\mu_j \in \mathbb{R}; \hspace{5pt} \sigma_j > 0; \hspace{5pt} \kappa_j > 0\}$ is the set of constraints when margins are VG, whereas $\overline{\mathcal{M}}_j = \{-\gamma_j < \beta_j < \gamma_j; \hspace{5pt} \delta_j > 0; \hspace{5pt} \gamma_j > 0\}$ is the set of constraints when margins are NIG.

In the above expression, $N_j$ is the number of market quotes used to fit the $j$-th margin and $ \omega_l, l=1,\ldots,N_j$ are arbitrary weights, fixed in advance.
The choice of the quotes and their weights can follow multiple criteria.
First, it is preferred to only use quotes of highly traded products; in these cases, such quotes can be considered accurate proxies of the fair values of these products, and provide reliable information about the log-return distributions of the underlying assets.
In addition, it can be convenient to assign larger weights to the quotes that are more relevant for our objective; if we are only interested in pricing, say, exotic products with short-term maturity, then we should calibrate our model mainly on short-term options.
This is computationally more efficient and in general improves the calibration performance.
A further discussion on this matter is reported in Section \ref{Calibration Choices}, in which we focus on our case study.

Once the optimized marginals $\{ \mathcal{M}^*_1, ..., \mathcal{M}^*_n \}$ are obtained, dependence parameters $\mathcal{D}$ are fitted by
solving
\begin{equation}\label{dependence-calibration}
\begin{aligned}
\mathcal{D}^*
=
\hspace{5pt}
& \underset{
\mathcal{D}
}{\text{argmin}}
& & \frac{2}{n (n-1)}
\sum_{i=1}^n
\sum_{j>i}
\big[ \rho_{\boldsymbol{Y}}^{\text{mod}} (i, j, \mathcal{M}^*_i, \mathcal{M}^*_j, \mathcal{D}) - \rho_{\boldsymbol{Y}}^{\text{mkt}} (i,j) \big]^2
\\
& \text{subject to}
& & \mathcal{D} \in \overline{\mathcal{D}}
\end{aligned}
\end{equation}
where $\overline{\mathcal{D}}$ is a model-specific feasible region to be specified in Subsection \ref{Model-Specific Calibrations}.
Also, assuming the availability of a single quoted price (for each pair of assets) from a liquidly-traded multi-name instrument, $\rho_{\boldsymbol{Y}}^{\text{mkt}} (i,j)$ is the Black-Scholes implied correlation from that quote.
Analogously, $\rho_{\boldsymbol{Y}}^{\text{mod}} (i, j, \mathcal{D})$ would be the implied correlation from the model price of the multi-name product.
Multiple correlations along strikes and maturities might be available, but it is more frequent to dispose of a restricted number of them, or even just one derived from a single quote estimated according to the beliefs of traders, rather than an actual liquid quote.
Another common situation is to have no quoted prices at all, in which case $\rho_{\boldsymbol{Y}}^{\text{mkt}} (i,j)$ is sometimes approximated by the historical correlation, and $\rho_{\boldsymbol{Y}}^{\text{mod}} (i, j, \mathcal{D})$ is approximated by the theoretical correlation (see, e.g., Eq.\ \eqref{rhoY-LS-VG}), abandoning the risk-neutral measure.

\subsection{Model-Specific Calibrations}\label{Model-Specific Calibrations}

Having presented the commonly used two-step calibration, we now examine alternative formulations that can help models to reduce the observed trade-off between marginal and dependence fit.
To this aim, we show tailor-made calibration settings of constrained LS and BB processes. Also, we specify the only possible fitting procedure for unconstrained models, that are supposed to reduce trade-off by construction.

In the constrained LS models, 
dependence parameters are  $\mathcal{D} = \left(a, \{ \rho_{ij} \}_{i \neq j}\right)$, while the feasible region is given by
$$
\overline{\mathcal{D}} = \Big\{ 0 < a < \min_j \Big( \kappa_j^{-m} \Big);
\hspace{5pt}
-1 \leq \rho_{ij} \leq 1, \forall i \neq j \Big\},
$$
where $m=1$ in the LS-VG case, and $m=0.5$ in LS-NIG.
As anticipated, the above bounds on $a$ lead to a potential trade-off between correlation range and marginal kurtosis fit.
As marginal parameters are calibrated before dependence ones, calibration error is likely to be concentrated on correlation.
To better distribute the error, \cite{marena2018pricing} propose to run a joint calibration instead, consisting of performing marginal fit of all the assets at the same time, while imposing a maximum level of correlation gap allowed, $\epsilon>0$.
In particular,
\begin{equation}\label{joint calibration}
\begin{aligned}
\{\mathcal{M}^*_1, ..., \mathcal{M}^*_n, \mathcal{D}^*\}
=
\hspace{5pt}
& \underset{
\mathcal{M}_1, ..., \mathcal{M}_n, \mathcal{D}
}{\text{argmin}}
& & \sum_{j = 1}^n
\frac{1}{N_j} \sum_{l=1}^{N_{j}} \big[ v_l^{\text{mod}} (\mathcal{M}_j) - v_l^{\text{mkt}} \big]^2
\\
& \text{subject to}
& & \mathcal{M}_1 \in \overline{\mathcal{M}}_1, ..., \mathcal{M}_n \in \overline{\mathcal{M}}_n, \mathcal{D} \in \overline{\mathcal{D}}
\\
&&& \abs{
\rho_{\boldsymbol{Y}}^{\text{mod}} (i, k, \mathcal{D}) - \rho^{\text{mkt}} (i,k)
}
< \epsilon
,
\hspace{10pt}
i \neq k.
\end{aligned}
\end{equation}

As it can be seen, Eq.\ \eqref{joint calibration} does not include dependence parameters in the objective function.
This might lead to situations where marginal kurtosis are already low, and we prevent calibration from finding a correlation error lower than $\epsilon$.
In practice, joint calibration is preferably run only after the (less expensive) two-stage calibration has failed to reach an acceptable correlation fit.
In such cases, the trade-off between marginals and correlation should push correlation error to get closer to $\epsilon$.
Another approach to deal with the mentioned trade-off can be to run the usual two-stage calibration, but adding a suitable upper bound on $\kappa_j$ to the feasible region $\overline{\mathcal{M}}_j$, $j=1...,n$.
This could avoid the need for a joint calibration, whose optimization problem could struggle with high dimensionality.
However, choosing such upper bound requires to have some previous knowledge of the market condition, which can be non-trivial.

Concerning the constrained BB-VG model,
the dependence parameters are given by $\mathcal{D} = (b_1, \ldots, b_n, \mu_Z, \sigma_Z, \kappa_Z)$ and the feasible region is
\begin{equation}\label{BB-VG feasible region}
\begin{aligned}    
\overline{\mathcal{D}} =
& \: \{ b_j \in \mathbb{R}, \forall j;\: \mu_Z \in \mathbb{R};\: \sigma_Z>0;\: \kappa_Z>0 \}
\hspace{1pt}
\\ & \cap \: \{ \sigma_j^2 - b_j \sigma_Z^2 >0, \forall j;\: \kappa_Z - \kappa_j >0, \forall j \}
\hspace{1pt}
\\ & \cap \: \{ \kappa_j \mu_j = \kappa_Z b_j \mu_Z, \forall j;\: \kappa_j \sigma^2_j = \kappa_Z b^2_j \sigma^2_Z, \forall j \}.
\end{aligned}
\end{equation}
In the constrained BB-NIG model,
$\mathcal{D} = (b_1, \ldots, b_n, \beta_Z, \delta_Z, \gamma_Z)$ are the dependence parameters, while the feasible region is
\begin{equation}\label{BB-NIG feasible region}
\begin{aligned}    
\overline{\mathcal{D}} =
& \: \{ b_j \in \mathbb{R}, \forall j;\: -\gamma_Z < \beta_Z < \gamma_Z;\: \delta_Z>0;\: \gamma_Z>0 \}
\hspace{1pt}
\\ & \cap \: \{ \delta_j - b_j \delta_Z >0, \forall j \}
\hspace{1pt}
\\ & \cap \: \{ \beta_j = b_j^{-1} \beta_Z, \forall j;\: \gamma_j = b_j^{-1} \gamma_Z, \forall j \}.
\end{aligned}
\end{equation}
The first and second sets of constraints of \eqref{BB-VG feasible region} and \eqref{BB-NIG feasible region} ensure that parameters of the systematic and the idiosyncratic components satisfy their domains.
The third sets, that we resume from \eqref{VG-conv-restrictions-1} and \eqref{NIG-conv-restrictions-1}, are needed to meet convolution conditions \eqref{conv-conditions}. \cite{ballotta2016multivariate} observed that such equality constraints are often satisfied only up to a least squares approximation.

In practice, dependence calibration of constrained BB models can be reformulated as a relaxed problem where restrictions \eqref{VG-conv-restrictions-1} or \eqref{NIG-conv-restrictions-1} are shifted to the objective function as follows:
\begin{equation}\label{BB dependence-calibration 2}
\begin{aligned}
\mathcal{D}^*
=
\hspace{5pt}
& \underset{
\mathcal{D}
}{\text{argmin}}
& & \frac{2}{n (n-1)}
\sum_{i=1}^n
\sum_{j>i}
\big[ \rho_{\boldsymbol{Y}}^{\text{mod}} (i, j, \mathcal{M}^*_i, \mathcal{M}^*_j, \mathcal{D}) - \rho_{\boldsymbol{Y}}^{\text{mkt}} (i,j) \big]^2
+
h \sum_{j=1}^n (c_{j,1}^2 + c_{j,2}^2)
\\
& \text{subj.\ to}
& & \mathcal{D} \in \overline{\mathcal{D}} \setminus \eqref{VG-conv-restrictions-1}, \eqref{NIG-conv-restrictions-1}
\end{aligned}
\end{equation}
where
$$
\begin{cases}
c_{j,1} = \kappa_j \mu_j - \kappa_Z b_j \mu_Z,\: c_{j,2} = \kappa_j \sigma_j^2 - \kappa_Z b_j^2 \sigma_Z^2
&
\text{for VG}
\\
c_{j,1} = \beta_j - b_j^{-1} \beta_Z,\: c_{j,2} = \gamma_j - b_j^{-1} \gamma_Z
&
\text{for NIG}
\end{cases}
$$
for some $h>0$.
An alternative approach that avoids the arbitrary parameter $h$ is to consider an objective function that only includes the convolution-related penalty term, while the correlation term is expressed as a constraint.
The optimization problem would then become
\begin{equation}\label{BB dependence-calibration}
\begin{aligned}
\mathcal{D}^*
=
\hspace{5pt}
& \underset{
\mathcal{D}
}{\text{argmin}}
& & \sum_{j=1}^n (c_{j,1}^2 + c_{j,2}^2)
\\
& \text{subj.\ to}
& & \mathcal{D} \in \overline{\mathcal{D}} \setminus \eqref{VG-conv-restrictions-1}, \eqref{NIG-conv-restrictions-1}
\\
&&& \abs{
\rho_{\boldsymbol{Y}}^{\text{mod}} (i, k, \mathcal{D}) - \rho^{\text{mkt}} (i,k)
}
< \epsilon
,
\hspace{10pt}
i \neq k.
\end{aligned}
\end{equation}
Although a trade-off between marginal and dependence fit is not apparent in BB theoretical construction, it is still possible to encounter it in practice, when we try to meet convolution conditions.
For this reason, either the penalty coefficient $h$ in \eqref{BB dependence-calibration 2} or the maximum correlation error allowed $\epsilon$ in \eqref{BB dependence-calibration} should be suitable tuned to shift calibration error on one side or the other.
The resulting convolution errors of \eqref{BB dependence-calibration 2} or \eqref{BB dependence-calibration} can be then measured as differences between marginal moments of the laws of $Y_j$ and $X_j + b_j Z$, $j=1, ..., n$, or comparing calibration fits of the these two distributions.

With regard to the unconstrained versions of LS and BB models, dependence parameters enter the marginal parametrization.
A two-stage calibration to fit these models is then improper, since each marginal calibration would subsequently change the value of a dependence parameter.
The only way to estimate these processes is by means of a joint calibration as in \eqref{joint calibration},
with the important difference that the feasible region $\overline{\mathcal{D}}$ only includes domain-related restrictions.

\section{Worst-Performance Derivatives}\label{Worst Performance Derivatives}

\begin{figure}
    \centering
    \includegraphics[width=8cm]{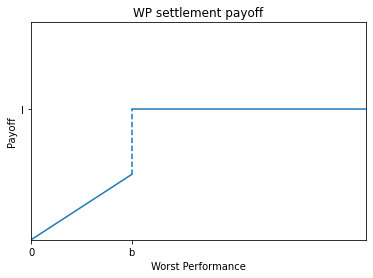}
    \caption{
    \small
    Payoff structure at settlement date of the three WP products, excluding the last coupon and a possible early redemption.
    It can be seen as a strategy involving a long position in a bond and a short one in a down-and-in put option.
    $I$ = issue price. $b$ = terminal barrier.
    }
    \label{fig:WP_payoff}
\end{figure}

Before testing the calibration and pricing skills of the models, we describe the exotic products to be priced in the analysis, whose payoffs are also relevant to make suitable calibration choices.
We consider three multi-asset derivatives issued by Intesa Sanpaolo bank: \textit{Standard Long Barrier Plus Worst Of Certificates}; \textit{Standard Long Barrier Digital Worst Of Certificates}; \textit{Standard Long Autocallable Barrier Digital Worst Of Certificates with Memory Effect}.
From now on we call them \textit{WP1}, \textit{WP2}, \textit{WP3}, respectively, as their payoffs depend on the \textit{worst performance} (WP) among underlyings. 
Also, they can be seen as a strategy involving a long position in a coupon bond and a short one in a down-and-in put option. Payoff details are described in what follows, while a graphical representation of the payoff at maturity is depicted in Fig.\ \ref{fig:WP_payoff}.

Let $\{t_0, ..., t_m\}$ be a set of relevant contract dates and $\mathbf{S} = (S_1, ..., S_n)$ be a vector of underlying assets, so that $\{\mathbf{S} (t)\}_{t \geq 0}$ represents the joint process of the assets. Let also the performance vector be defined as
$$
\mathbf{P} (t_i) = (P_1(t_i), ..., P_n(t_i)) = \Bigg( \frac{ S_1(t_i) }{ S_1(t_0) }, ..., \frac{S_n(t_i)}{ S_n(t_0)} \Bigg)
,
\hspace{15pt}
i = 1, ..., m,
$$
along with the associated worst-performance assets
$$
w_i = \underset{j \in \{1, ..., n\} }{ \text{argmin}} \{P_j (t_i)\}
,
\hspace{15pt}
i = 1, ..., m
.
$$
Define also the barrier events as
\begin{equation*}
\begin{aligned}   
&E = \{ P_{w_m}(t_m) < b \},
\\
&E_{d,i} = \{ P_{w_i}(t_i) < b_{d,i} \},
&i = 1, ..., m
\\
&E_{r,i} = \{ P_{w_i}(t_i) > b_{r,i} \},
&i = 1, ..., m
\end{aligned}
\end{equation*}
where $b, b_{d,1}, ..., b_{d,m}, b_{r,1}, ..., b_{r,m}$ are predefined barriers (in percentages).
Denoting by $I$ the issue price and by $k$ the periodic coupon payment, we get WP1 and WP2 payoffs respectively as
\begin{equation*}
\begin{aligned}
&
\Pi_{\text{WP1}} = k m + \pi_m
\\
&
\Pi_{\text{WP2}} = k \sum_{i = 1}^m \mathbbm{1}_{E_{d,i}^c} + \pi_m
\end{aligned}
\end{equation*}
where
$$
\pi_m =
I \: \mathbbm{1}_{E^c}
+
\frac{S_{w_m} (t_m) }{ S_{w_m} (t_0)} I \: \mathbbm{1}_E
,
$$
and the superscript $c$ denotes the complement event.
Note that the payoff presented here assumes we are pricing the derivative at the issuing date, while it is often the case we buy it at a future date, missing some intermediate earnings (e.g., coupons). 

WP3 keeps a similar structure as WP2, but it adds a \textit{memory effect} and \textit{redemption} features.
The first means that at each date $i$, provided that the worst performance is above the barrier $b_{d,i}$, the investor gains the current coupon plus the previous coupons that have not been paid.
The second implies that at the first date on which the event $E_{r,i}$ is verified, if it does happen, the investor receives the issue price and the contract expires. For sake of completeness, we report the algorithm needed to compute the discounted WP3  payoff for a single Monte Carlo trajectory in Appendix \ref{app:WP3 algorithm}.

\section{Empirical Analysis}\label{Empirical Analysis}

At this point, we have all the elements to perform an empirical test of the analyzed L\'evy models, with the aim of evaluating the ability of such models to fit market data and produce a fair valuation of the exotic derivatives.


We start this section by describing our dataset. 
Regarding calibration data, as it is generally the case, quotes of liquid multivariate instruments are not available. As a consequence, calibration data includes two distinct parts.
The first consists of liquid implied volatility surfaces of single-asset options, needed to fit marginal laws. 
The second involves an estimate of market correlation between log-returns of the assets. For each pair of assets, only a single market implied correlation is available, that is assumed to be constant over maturities and moneyness. Also, such correlation is implied by an ideal price estimated by traders, rather than the market price of a liquidly traded product.
We consider such calibration data for 13 monthly-distanced calibration dates (from 30/11/2021 to 30/11/2022), in order to assess model performance under various market conditions.

Concerning pricing data, the actively-traded WP derivatives that we price with L\'evy models have the following characteristics, besides the payoff features described in Section \ref{Worst Performance Derivatives}:
\begin{itemize}
    \item  WP1 is written on EURO STOXX 50 and FTSE MIB indexes, it has been issued in 28/03/2019 and expires in 28/03/2023.
    \item WP2 is written on EURO STOXX Select Dividend 30 and FTSE MIB indexes, it has been issued in 30/03/2020 and expires in 30/03/2026.
    \item WP3 is written on RWE AG, \'Electricit\'e de France SA and Iberdrola SA shares, it has been issued in 16/11/2021 and expires in 18/11/2024.
\end{itemize}
For these instruments, market bid prices are available at all dates where models are calibrated, even if the liquidity of WP products is relatively scarce.
As a consequence, it is possible to compare WP prices computed by the models with benchmarks provided by market data.

\subsection{Calibration Choices}\label{Calibration Choices}

As a starting point for our calibration analysis, we report some preliminary decisions on the adopted strategies to fit models.
Depending on the specific cases, such choices derive from theoretical considerations and/or a \textit{first run} of calibration and pricing, useful to get a flavour of the dataset.


Concerning the optimization techniques employed in the calibration problem, we decide to use a stochastic search algorithm to fit model parameters, as done for example by \cite{cont2004recovering}.
Specifically, we use a versatile Differential Evolution algorithm \citep{storn1997differential}, available in \textit{scipy} Python library.
We then refine the solution provided by the pure DE by employing a local optimizer.
Differently, other authors (e.g., \citealp{cont2004nonparametric} and \citealp{alfeus2020regularization}) employ multi-start methods for the calibration problems.
As these algorithms exploit the gradient of the objective function, they can often converge to solutions relatively quickly.
However, as non-Gaussian L\'evy distributions are likely to display a highly non-convex objective function (see VG case in \citealp{cont2004nonparametric}), using a multi-start algorithm risks to excessively depend on the grid choice.
Another approach adopted by researchers is to fit models via surrogate-like optimization (see \citealp{ruf2020neural} for a literature review of Artificial Neural Networks for pricing applications).
These techniques are normally used when objective function is computationally expensive.
However, in our case the objective function only consists, for marginals, of computing a few prices via Fourier techniques (we use the COS method of \citealp{fang2009novel}) and their implied volatilities.
Still, ANNs have the benefit to shift computational burden \textit{offline}, but this can be more desired in a context of frequent recalibrations, while our study just involves a few time-distanced calibrations.
Similarly, we decide not to employ regularization techniques (see e.g., \citealp{cont2006retrieving},
\citealp{egger2005tikhonov}, \citealp{crepey2010tikhonov}, and \citealp{dai2016calibration}), as these are more needed in presence of frequent recalibrations.
To be surer that regularization can be avoided, we apply small data perturbations to a first run of calibration, finding out that results are stable enough to such perturbations.

A second important decision regards the points of the volatility surfaces to use for model fitting.
As WP payoffs are largely linked to the underlying asset prices at settlement date, it can be convenient to assign larger weights (see Eq.\ \eqref{marginal calibration}) to calibration points close to WP settlement dates.
In fact, as L\'evy processes have i.i.d.\ increments, we cannot expect them to perfectly fit whole volatility surfaces, as such surfaces reflect the well-known non-stationarity of realized log-returns. As a consequence, concentrating calibration effort on a lower number of volatility surface points seems reasonable.
After these preliminary considerations and looking at a first run of calibration and pricing, we actually notice that the best strategy is to just calibrate models on 2-3 volatility smiles close to WP settlement dates. Indeed, although intermediate contract payments risk to be poorly predicted by the so-calibrated processes, we evidence that the overall pricing performance of the models benefit from our strategy.

Another calibration choice lies in the context of dependence fit.
That is, we decide to use \textit{theoretical} correlation (see, for example, Eq.\ \eqref{rhoY-LS-VG} in the case of constrained LS-VG) as a proxy of \textit{model-implied} correlation (i.e., Black-Scholes implied correlation from model price).
In fact, as market correlation is implied from an estimated market price (see the introduction of Section \ref{Empirical Analysis}), the most natural calibration choice would be to employ model-implied correlation to fit dependence structure.
However, we notice that using either model-implied or theoretical correlations yields nearly the same exotic prices produced by the calibrated models.
Therefore, as the computational time required to calculate theoretical correlation is substantially lower than that of model-implied correlation, we use the former.

To conclude the set of preliminary choices, we report our decisions concerning the calibration methods employed to fit the specific L\'evy models, discussed in the introduction of Section \ref{Calibration Methods} and in Subsection \ref{Model-Specific Calibrations}.
Regarding constrained LS models, as two-step calibration is normally faster than joint calibration, we first employ the former, and only when trade-off between marginal and dependence fit is strong, we perform a joint calibration to distribute calibration errors proportionally (similarly to \citealp{marena2018pricing}).
For constrained BB models, as their correlation coefficients do not make the trade-off clear from a theoretical perspective, we first set a tiny maximum correlation error allowed (see Eq.\ \eqref{BB dependence-calibration} and the discussion below), and when this strategy leads to excessively large convolution errors, we let correlation error increase.
Finally, as the theoretical structure of unconstrained models requires joint calibrations as in Eq.\ \eqref{joint calibration}, for this models we always follow the joint procedure, initially imposing a low bound on correlation error and increasing it if necessary.

\subsection{Calibration Results}\label{Calibration}

Having introduced the calibration setting, we are now ready to report results.
In order to show the variety of market conditions encountered in the analysis and to provide useful benchmarks to future researchers, we illustrate some statistics of the calibrated marginal parameters of constrained models in Table \ref{tab:Marginal Parameters}.
Such parameters belong to well-known univariate laws such as Variance Gamma and Normal Inverse Gaussian, which are embedded by both LS and BB constrained models, as such models preserve margins of known classes by construction.
Note though that BB processes are subject to convolution errors in the second stage of the two-step calibration, so parameters of Table \ref{tab:Marginal Parameters} only approximately represent BB marginal distributions.

Going into detail, Table \ref{tab:Marginal Parameters} reports the minimum, mean and maximum values of marginal parameters over the 13 calibration dates of the analysis.
A relevant point comes from observing that minima (or maxima) are sometimes similar across underlying assets.
This happens because we arbitrarily restrict parameters to take values in reasonably narrow ranges.
According to the aforementioned first run of calibration, we notice that such an approach is useful for at least a couple of reasons.
First, and not surprisingly, it increases the stability and convergence speed of the calibration algorithm.
Secondly, it often reduces the overall calibration error, as correlation and convolution fit are likely to suffer from too small, or too large, values of marginal parameters.
Although tightening parameter bounds might seem a strong limitation to marginal fit, the non-identifiability of parameters, typical of non-Gaussian L\'evy distributions, is likely to help models to still find combinations of parameters that well-reflect market data.

\begin{table}[]
\scriptsize
\centering
\begin{tabular}{|l|c|c|c|c|c|c|c|c|c|}
\hline
& \multicolumn{3}{|c|}{$\mu$} & \multicolumn{3}{|c|}{$\sigma$} & \multicolumn{3}{|c|}{$\kappa$}
\\
\hline
Underlying & min & mean & max & min & mean & max & min & mean & max
\\
\hline
EURO STOXX 50 &
-0.3062 & -0.1922 & -0.1223 & 0.1665 & 0.203 & 0.2196 & 0.7068 & 1.7768 & 2.997
\\
FTSE MIB (a) &
-0.3326 & -0.2206 & -0.1318 & 0.1669 & 0.2238 & 0.2586 & 0.4452 & 1.562 & 2.9988
\\
EURO STOXX S.D. 30 &
-0.1559 & -0.1484 & -0.1364 & 0.0061 & 0.0628 & 0.1602 & 2.9917 & 2.9986 & 2.9998
\\
FTSE MIB (b) &
-0.1902 & -0.1706 & -0.1504 & 0.0059 & 0.0525 & 0.1866 & 2.997 & 2.9991 & 3.0
\\
\'ELEC.\ DE FRANCE SA &
-1.0136 & -0.2551 & -0.1782 & 0.1204 & 0.2984 & 0.3545 & 0.0865 & 0.6599 & 1.7835
\\
IBERDROLA SA &
-0.1929 & -0.1538 & -0.1299 & 0.1898 & 0.2184 & 0.2457 & 0.8713 & 1.5145 & 2.592
\\
RWE AG &
-0.8346 & -0.1337 & -0.0798 & 0.0055 & 0.3155 & 0.3865 & 0.1194 & 1.874 & 2.9927
\\
\hline
\hline
& \multicolumn{3}{|c|}{$\beta$} & \multicolumn{3}{|c|}{$\delta$} & \multicolumn{3}{|c|}{$\gamma$}
\\
\hline
Underlying & min & mean & max & min & mean & max & min & mean & max
\\
\hline
EURO STOXX 50 &
-4.9999 & -4.7559 & -3.7211 & 0.1 & 0.1594 & 0.2359 & 4.7087 & 6.1767 & 6.6663
\\
FTSE MIB (a) &
-4.9998 & -4.346 & -2.3517 & 0.1013 & 0.1934 & 0.322 & 3.0398 & 5.8379 & 6.6664
\\
EURO STOXX S.D. 30 &
-4.6967 & -4.1135 & -3.6941 & 0.1 & 0.1 & 0.1001 & 4.7693 & 5.2626 & 5.7746
\\
FTSE MIB (b) &
-3.1203 & -2.7794 & -2.1365 & 0.1 & 0.102 & 0.1095 & 2.9566 & 3.6037 & 3.9245
\\
\'ELEC.\ DE FRANCE SA &
-3.335 & -2.2754 & -1.4409 & 0.2406 & 0.3686 & 0.4997 & 2.9628 & 4.4852 & 5.9827
\\
IBERDROLA SA &
-4.9961 & -3.5964 & -2.2539 & 0.1282 & 0.1759 & 0.2221 & 3.4076 & 5.3033 & 6.6655
\\
RWE AG &
-2.7009 & -1.0942 & -0.7129 & 0.1616 & 0.2203 & 0.2812 & 1.6059 & 2.5039 & 3.7091
\\
\hline
\end{tabular}
\caption{
\small
VG and NIG marginal parameters of underlying assets. We report their minimum, average and maximum over the 13 calibration dates.
}
\label{tab:Marginal Parameters}
\end{table}

\begin{figure} 
\centering
\subfloat[Marginal calibration errors of WP1 underlyings.
\label{VG marginal calibration errors WP1}
]{\includegraphics[scale=0.5]{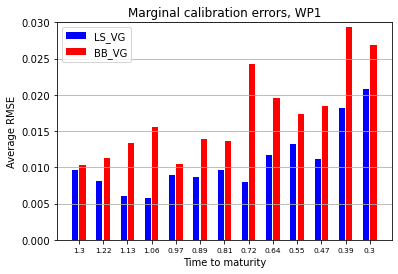}}
\hfill
\subfloat[Dependence calibration errors of WP1 underlyings.
\label{VG dependence calibration errors WP1}
]{\includegraphics[scale=0.5]{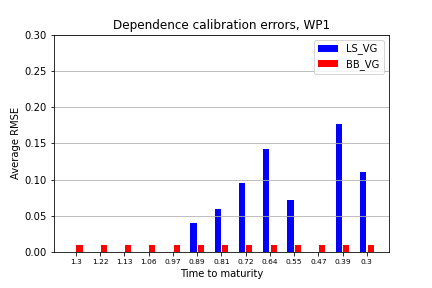}}
\newline
\subfloat[Marginal calibration errors of WP2 underlyings.
\label{VG marginal calibration errors WP2}
]{\includegraphics[scale=0.5]{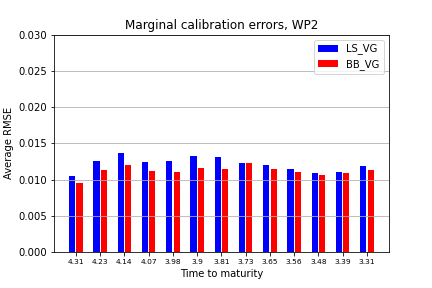}}
\hfill
\subfloat[Dependence calibration errors of WP2 underlyings.
\label{VG dependence calibration errors WP2}
]{\includegraphics[scale=0.5]{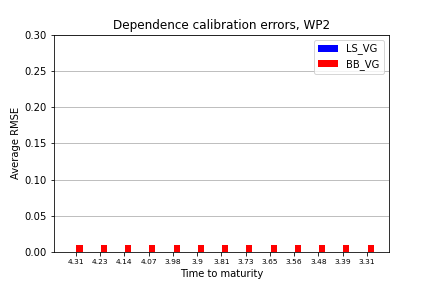}}
\newline
\subfloat[Marginal calibration errors of WP3 underlyings.
\label{VG marginal calibration errors WP3}
]{\includegraphics[scale=0.5]{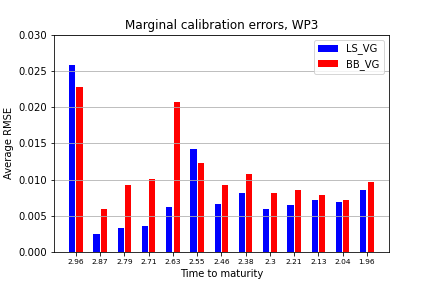}}
\hfill
\subfloat[Dependence calibration errors of WP3 underlyings.
\label{VG dependence calibration errors WP3}
]{\includegraphics[scale=0.5]{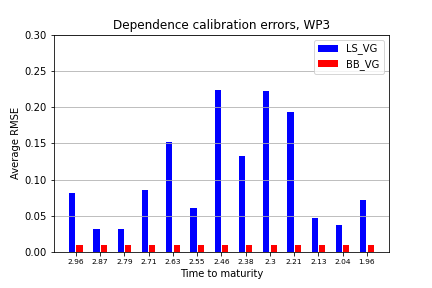}}
\caption{
\small
Constrained LS-VG and BB-VG calibration errors for the 13 considered times to maturity. Left-hand side figures represent average RMSEs between model and market volatilities of the basket of underlyings of each WP. Right-hand side figures represent average RMSEs between model and market correlations of the set of underlying pairs of each WP.
}
\label{fig:VG calibration errors}
\end{figure}

\begin{figure} 
\centering
\subfloat[Marginal calibration errors of WP1 underlyings.
\label{NIG marginal calibration errors WP1}
]{\includegraphics[scale=0.5]{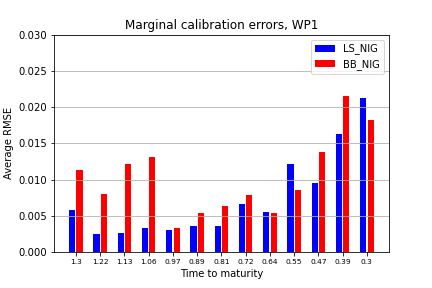}}
\hfill
\subfloat[Dependence calibration errors of WP1 underlyings.
\label{NIG dependence calibration errors WP1}
]{\includegraphics[scale=0.5]{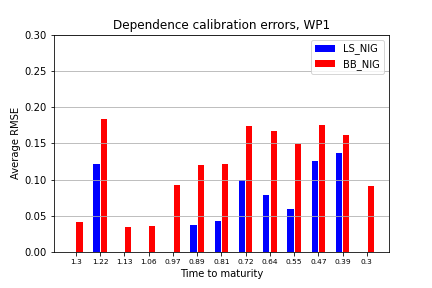}}
\newline
\subfloat[Marginal calibration errors of WP2 underlyings.
\label{NIG marginal calibration errors WP2}
]{\includegraphics[scale=0.5]{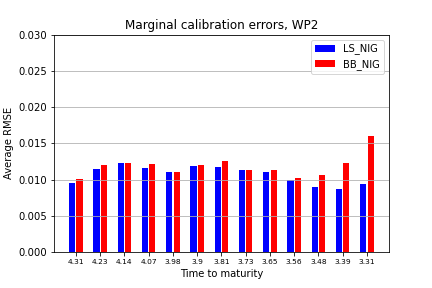}}
\hfill
\subfloat[Dependence calibration errors of WP2 underlyings.
\label{NIG dependence calibration errors WP2}
]{\includegraphics[scale=0.5]{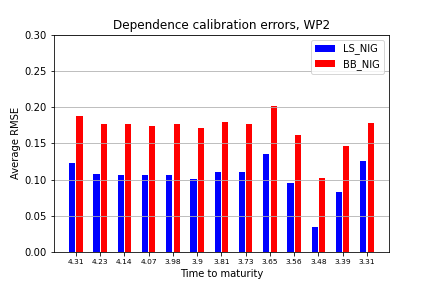}}
\newline
\subfloat[Marginal calibration errors of WP3 underlyings.
\label{NIG marginal calibration errors WP3}
]{\includegraphics[scale=0.5]{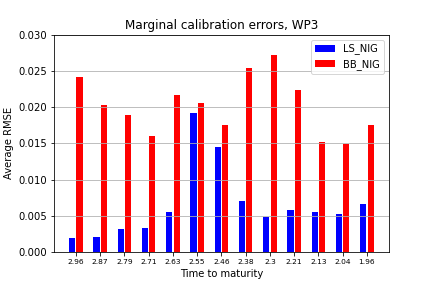}}
\hfill
\subfloat[Dependence calibration errors of WP3 underlyings.
\label{NIG dependence calibration errors WP3}
]{\includegraphics[scale=0.5]{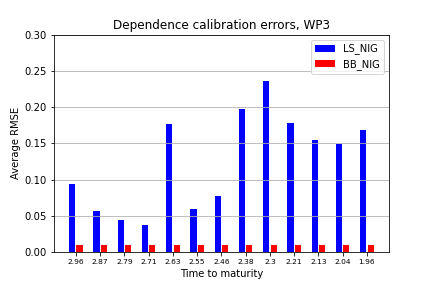}}
\caption{
\small
Constrained LS-NIG and BB-NIG calibration errors for the 13 considered times to maturity. Left-hand side figures represent average RMSEs between model and market volatilities of the basket of underlyings of each WP. Right-hand side figures represent average RMSEs between model and market correlations of the set of underlying pairs of each WP.
}
\label{fig:NIG calibration errors}
\end{figure}

After displaying the statistics of marginal parameters, we can now present the calibration errors of constrained LS and BB models, in order to evaluate their fitting performances and explore a possible trade-off between marginal and dependence structure.
We display the average root mean square error (RMSE) between model and market implied volatilities of WP underlying assets in the left-hand side of Figs.\ \ref{fig:VG calibration errors} and \ref{fig:NIG calibration errors}, and between model and market correlation, in the right-hand side.
In our first calibration and pricing run, we notice that the effect of correlation error on the exotic price is of the order of magnitude of about one tenth of volatility error.
As a consequence, we keep this ratio of scales between marginal and dependence plots of Figs.\ \ref{fig:VG calibration errors} and \ref{fig:NIG calibration errors}, so that the reader can assign nearly the same importance to all error bars.

According to Figs.\ \ref{VG marginal calibration errors WP1}, \ref{VG marginal calibration errors WP2} and \ref{VG marginal calibration errors WP3}, the marginal calibration of multivariate VG models normally yields lower errors for LS-VG than for BB-VG.
This is expected, as BB adds convolution errors to the marginal calibration errors outputted by the first part of the two-step calibration (note, however, that in some cases convolution error could accidentally improve marginal fit).
However, we notice in Figs.\ \ref{VG dependence calibration errors WP1} and \ref{VG dependence calibration errors WP3} that the better marginal calibration of LS-VG comes at the expense of correlation fit, that become significant in a few cases, remarking the presence of fitting trade-off.
On the other hand, the exceptionally good dependence calibration of BB-VG, coming from setting a small $\epsilon=0.01$ in the optimization problem of Eq.\ \eqref{BB dependence-calibration}, seems to offset the excess marginal error in most cases.
To summarize, while the results on WP1- and WP2-linked calibrations are somewhat ambiguous, BB-VG performs better than LS-VG in WP3 case.

Concerning calibration results of multivariate NIG models, some considerations can be made by observing Fig.\ \ref{fig:NIG calibration errors}.
Similarly to VG case, we detect a poorer marginal fit of BB-NIG with respect to LS-NIG. This is particularly evident in the WP3-linked marginal calibration errors of Fig.\ \ref{NIG marginal calibration errors WP3}, where we still use a calibration method that imposes a maximum correlation error allowed of 0.01.
In WP1- and WP2-linked calibrations, in order to avoid huge marginal errors, we are forced to relax the constraint on correlation fit.
The effect of such decision is clear from Figs.\ \ref{NIG dependence calibration errors WP1} and \ref{NIG dependence calibration errors WP2}, where BB dependence calibration errors get considerably larger than the other cases.
To sum up, we observe an overall better fit of LS-NIG model with respect to BB-NIG. Also, while LS-VG and LS-NIG enjoy relatively similar results among each other, we evidence a significant difference between BB-VG and BB-NIG fitting performances.
Such outcome is likely to be due to the sensitivity of BB model construction to the marginal law assigned to the process.
In fact, as can be seen from Eqs.\ \eqref{VG-conv-restrictions-1}, \eqref{VG-conv-restrictions-2}, \eqref{NIG-conv-restrictions-1} and \eqref{NIG-conv-restrictions-2}, the convolution conditions needed to preserve marginal laws can be quite different between VG and NIG cases, so is the ability of BB models to satisfy such conditions.

An important and more general consideration is that a trade-off between marginal and dependence fit exists in practice and can even be strong in some cases.
As trade-off is, at least in part, produced by the convolution conditions of constrained LS and BB models (recall, for example, the bounds on correlation parameter $a$ in Eqs.\ \eqref{Gamma-Sub} and \eqref{IG-Sub}), we then decide to test unconstrained versions of LS and BB models in Subsection \ref{Constrained vs Unconstrained Margins}, to find out whether they can significantly improve calibration fit.

\subsection{Pricing}\label{Pricing}

\begin{figure} 
\centering
\subfloat[WP1 prices.
\label{fig:prices 1}
]{\includegraphics[scale=0.5]{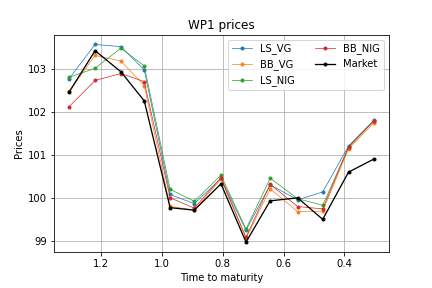}}
\hfill
\subfloat[WP1 percentage differences.
\label{fig:percentage differences 1}
]{\includegraphics[scale=0.5]{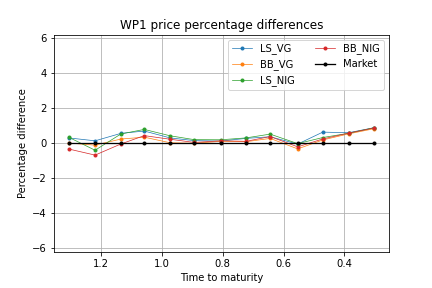}}
\newline
\subfloat[WP2 prices.
\label{fig:prices 2}
]{\includegraphics[scale=0.5]{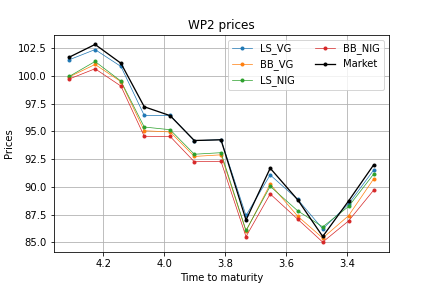}}
\hfill
\subfloat[WP2 percentage differences.
\label{fig:percentage differences 2}
]{\includegraphics[scale=0.5]{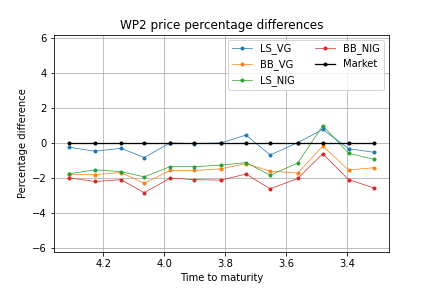}}
\newline
\subfloat[WP3 prices.
\label{fig:prices 3}
]{\includegraphics[scale=0.5]{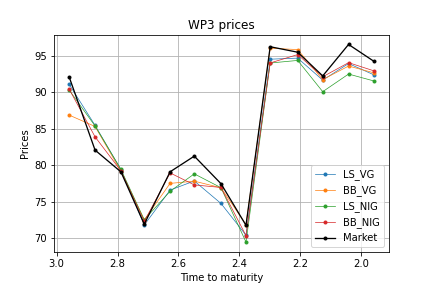}}
\hfill
\subfloat[WP3 percentage differences.
\label{fig:percentage differences 3}
]{\includegraphics[scale=0.5]{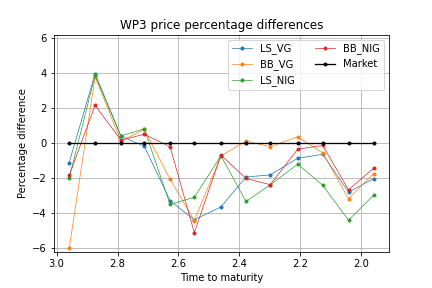}}
\caption{
\small
Prices and percentage differences between model and market prices, on the 13 considered times to maturity. Prices are expressed as percentage of the issue price. Percentage differences are computed as (Price $-$ Market Price) / Market Price $\times$ 100.
}
\label{fig:prices}
\end{figure}

Having calibrated the constrained models, we are now ready to price WP instruments, in order to investigate the ability of the models to produce a fair valuation of such derivatives.
Because of the exotic features of WP contracts, valuations are carried out via Monte Carlo methods, simulating enough sample paths to get suitably small MC errors.
As previously said, for sake of completeness we report the MC algorithm needed to price WP3 in Appendix \ref{app:WP3 algorithm}.
As market bid quotes of WP products are available, we display them together with model prices in the left-hand side of Fig.\ \ref{fig:prices}.
The right-hand side shows instead the percentage differences between model and market prices.

As a first level of comparison, we look at the differences between L\'evy prices.
It is clear from the plots that models enjoy comparable results among each other.
Also, percentage differences of the four models seem to follow a similar path over valuation dates, evidencing a homogeneity between model results.
However, an observation concerns WP2 pricing (see Figs.\ \ref{fig:prices 2} and \ref{fig:percentage differences 2}), where LS-VG replicates benchmarks particularly well, while BB-NIG produce the worst results among models.
Such evidence is, in fact, consistent with the results shown in the calibration plots (Figs.\ \ref{fig:VG calibration errors} and \ref{fig:NIG calibration errors}) and the related discussion, remarking the link between calibration and pricing.
Overall, although LS and BB models have different theoretical constructions, they preserve a similar ability to price exotics.

On the other hand, clearer evidences can be captured by comparing L\'evy models as a whole against market quotes.
Apart from short-maturity cases, we observe a tendency of the models to underestimate bid prices.
This is partly due to the calibrated model correlation not reaching the market level, that leads to a higher value of the embedded put option of the contract (recall WP payoff structure from Fig.\ \ref{fig:WP_payoff}), and so a lower overall price.
Whenever correlation enjoys a good fit, the overestimate of the put does not completely disappear.
This is possibly due to L\'evy distributions bearing fatter tails than those assumed by valuation models that concur in forging benchmark quotes.
In this regard, we remark that as WP products are not exchange-traded, their benchmark prices can significantly differ from their fair values; thus, the differences between model prices and benchmarks are not necessarily given by a poor valuation performance of the L\'evy models.

Looking in detail at the specific contracts, models enjoy the best results when pricing WP1 (see Figs.\ \ref{fig:prices 1} and \ref{fig:percentage differences 1}), where market bid quotes are replicated quite accurately, even with a slight and desirable high bias. 
Such good performance is consistent with the well-known ability of jump processes to price short-term options, even far from at-the-money positions.
Concerning the other contracts, WP2 is priced at longer times to maturity (see Figs.\ \ref{fig:prices 2} and \ref{fig:percentage differences 2}), making the correlation mismatch more effective.
As a result, percentage (absolute) differences increase. 
Also, differently from WP1, WP2 payoff contains some path-dependency (see Section \ref{Worst Performance Derivatives}), which is another source of inaccuracy if we price via stationary processes such as L\'evy ones (as observed, for example, by \citealp{ballotta2016multivariate}).
Regarding WP3 (Figs.\ \ref{fig:prices 3} and \ref{fig:percentage differences 3}), model and market prices diverge more for all models, for at least two reasons.
First, the exotic features of the contract increase substantially (as can be observed from the pricing algorithm in Appendix \ref{alg:WP3-payoff}).
Second, WP3 underlyings are stocks, with more irregular and less liquid volatility surfaces with respect to the underlying indexes of WP1 and WP2.
As a consequence, the link between calibration data and exotic quotes is more prone to mismatches.

Overall, except some pricing inaccuracy coming from the path-dependent features of the exotics, a significant source of error is likely to be calibration fit.
Especially if we look at the connection between LS-VG good pricing performance of WP2 product (Fig.\ \ref{fig:percentage differences 2}) and its relatively accurate fit (comparing all WP2-related calibrations in Figs.\ \ref{VG marginal calibration errors WP2}, \ref{VG dependence calibration errors WP2}, \ref{NIG marginal calibration errors WP2} and \ref{NIG dependence calibration errors WP2}), it seems logic to try to improve model fitting to produce more accurate valuations. This again points to test unconstrained models in Subsection \ref{Constrained vs Unconstrained Margins}, and check whether these models can better replicate market quotes.


\subsection{Sensitivities}\label{Sensitivities}

\begin{figure}[hbt!] 
\centering
\subfloat[Correlation effect on WP1 price for time to maturity 1.0 and moneyness 2.0, 1.6, 1.2.
\label{fig:correlation-effect 1}
]{\includegraphics[scale=0.5]{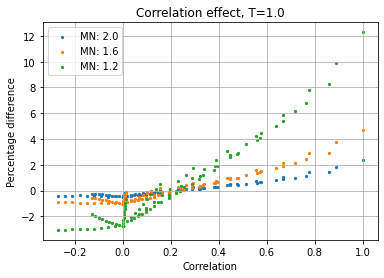}}
\hfill
\subfloat[Correlation effect on WP1 price for time to maturity 2.5 and moneyness 2.0, 1.6, 1.2.
\label{fig:correlation-effect 2}
]{\includegraphics[scale=0.5]{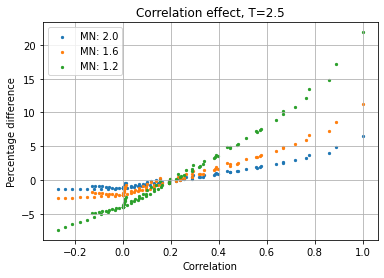}}
\newline
\subfloat[Correlation effect on WP1 price for time to maturity 4.0 and moneyness 2.0, 1.6, 1.2.
\label{fig:correlation-effect 3}
]{\includegraphics[scale=0.5]{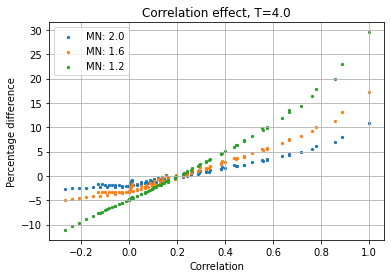}}
\hfill
\subfloat[Correlation effect on WP1 price for time to maturity 4.0 and moneyness 2.0.
\label{fig:correlation-effect 4}
]{\includegraphics[scale=0.5]{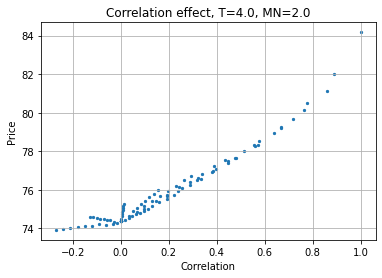}}
\caption{
\small
Correlation effect on WP1 price, using constrained LS-VG model and fixed marginal distributions $\mu_j=-0.15, \sigma_j=0.25, \kappa_j=1.60, \: j=1,...,n$. In Figs.\ (a)-(c),
each point represents the percentage difference between a price - with a certain combination of correlation and moneyness - and the average price between those with the same moneyness,
as a function of $\rho_{\boldsymbol{Y}}$. Fig.\ (d) shows prices as a function of $\rho_{\boldsymbol{Y}}$ in the realistic case where T=4.0 and MN=2.0, that is where WP1 is valued close to issue date. T = time to maturity. MN = moneyness (initial underlying price divided by strike).
}
\label{fig:correlation-effect}
\end{figure}

In order to have a better understanding of WP payoffs and to get the importance of making an accurate dependence calibration, we run a sensitivity-to-correlation analysis, by observing correlation effect on price in Fig.\ \ref{fig:correlation-effect}. 
As exotic products have comparable payoffs, we restrict the analysis to WP1 instrument.
Also, we just consider constrained LS-VG model, since models enjoy mutually similar sensitivities.
Furthermore, we fix marginal parameters to $\mu_j=-0.15,\: \sigma_j=0.25,\: \kappa_j=1.60,\: j=1,2$, representing a commonly-observed market situation according to Table \ref{tab:Marginal Parameters}.
Looking at Figs.\ \ref{fig:correlation-effect 1}, \ref{fig:correlation-effect 2}, \ref{fig:correlation-effect 3}, as expected from the payoffs of WP products, the price sensitivity to correlation increases for higher maturities and when moneyness gets closer to one. 
Moreover, in all instances sensitivity rises for very high correlation levels (commonly observed in the index market), evidencing a nonlinear effect on price and the importance of using models and calibration methods able to capture wide dependence ranges.
On a general level, it is clear from the plots that correlation has a large effect on WP1 price, remarking the need to distribute calibration errors in a proportionate way between marginal and correlation parameters.

\begin{figure}[hbt!] 
\centering
\subfloat[Nonlinear dep.\ effect on WP1 price for moneyness 1.6, correlation 0.6, 0.7, 0.8, 0.9.
\label{fig:nonlinear-dependence-effect 1}
]{\includegraphics[scale=0.5]{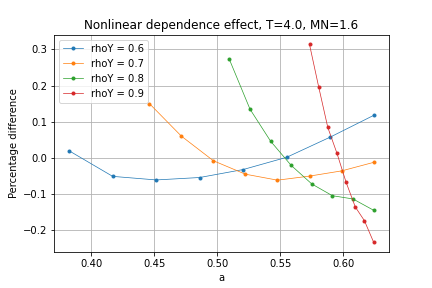}}
\hfill
\subfloat[Nonlinear dep.\ effect on WP1 price for moneyness 1.4, correlation 0.6, 0.7, 0.8, 0.9.
\label{fig:nonlinear-dependence-effect 2}
]{\includegraphics[scale=0.5]{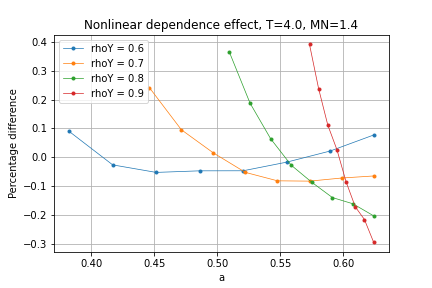}}
\newline
\subfloat[Nonlinear dep.\ effect on WP1 price for moneyness 1.2, correlation 0.6, 0.7, 0.8, 0.9.
\label{fig:nonlinear-dependence-effect 3}
]{\includegraphics[scale=0.5]{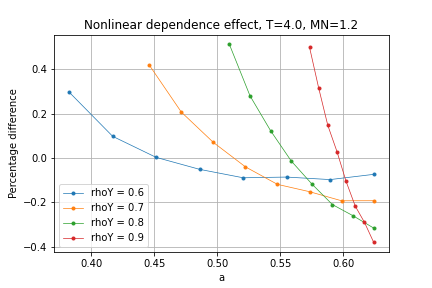}}
\hfill
\subfloat[Nonlinear dep.\ effect on WP1 price for moneyness 1.2, correlation 0.9.
\label{fig:nonlinear-dependence-effect 4}
]{\includegraphics[scale=0.5]{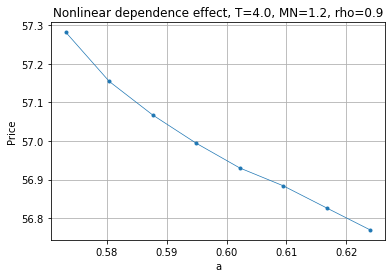}}
\hspace{0.6cm}
\caption{
\small
Nonlinear dependence effect on WP1 price, using constrained LS-VG model and fixed marginal distributions $\mu_j=-0.15, \sigma_j=0.25, \kappa_j=1.60, \: j=1,...,n$.
In Figs.\ (a)-(c),
each point represents the percentage difference between a price - with a certain combination of correlation and moneyness - and the average price between those with the same moneyness,
as a function of $a$. In Fig.\ (d) we plot prices as a function of $a$ for one of the considered instances, where price is relatively sensitive to nonlinear dependence. T = time to maturity. MN = moneyness (initial underlying price divided by strike).
}
\label{fig:nonlinear-dependence-effect}
\end{figure}

To explore the effect of nonlinear dependence on the valuation of exotics, we examine the price sensitivity to such feature in Fig.\ \ref{fig:nonlinear-dependence-effect},
recalling that it can only be observed in LS models. 
Again, we just consider 
WP1 product, constrained LS-VG model and fixed marginal laws with $\mu_j=-0.15,\: \sigma_j=0.25,\: \kappa_j=1.60, \: j=1,2$.
We observe the effect of nonlinear dependence by computing WP1 price for multiple values of parameter $a$
(see e.g., 
Eq.\ \eqref{Gamma-Sub}).
$a$ is the only parameter that affects nonlinear dependence in constrained LS models, but it also drives part of the correlation.
As a consequence, in order to isolate the effect of nonlinear dependence, we move $a$ while keeping $\rho_{\boldsymbol{Y}}$ (from Eq.\ \eqref{rhoY-LS-VG}) constant, by changing accordingly the other correlation drivers $\rho_{ij}, \forall i \neq j$.
The calculation is repeated for different correlation values, and for different levels of moneyness, as shown in Figs.\ \ref{fig:nonlinear-dependence-effect 1}, \ref{fig:nonlinear-dependence-effect 2}, \ref{fig:nonlinear-dependence-effect 3}.
We observe that when underlying asset prices are far from the barrier (Fig.\ \ref{fig:nonlinear-dependence-effect 1}),
there is a negative relationship between $a$ and prices when $\rho_{\boldsymbol{Y}}$ is very high, while relationship flattens and even become positive as we progress towards lower correlations.
As moneyness decreases (Fig.\ \ref{fig:nonlinear-dependence-effect 3}), we observe a clearer negative dependence for each level of $\rho_{\boldsymbol{Y}}$.
As a first point, we then establish that there exists some link between nonlinear dependence and WP1 price.
However, as $\rho_{\boldsymbol{Y}}$ increases, we note that the domain of $a$ shrinks, because we cannot set high correlations without a high $a$.
This eliminates the possibility to observe the whole relationship between nonlinear dependence and price.
In general, as expected, the order of magnitude of the effect of nonlinear dependence on price is much lower with respect to that of correlation, as can be seen comparing Figs.\ \ref{fig:correlation-effect 4} and \ref{fig:nonlinear-dependence-effect 4}.
On the one hand, the visible link between nonlinear dependence and price rewards LS models for their ability to capture such feature. Also, the small admissible range of parameter $a$ encourages to try to construct models that can accommodate larger nonlinear dependence ranges.

\subsection{Constrained vs Unconstrained Margins}\label{Constrained vs Unconstrained Margins}

Having shown results for constrained LS and BB models, it is now time to calibrate unconstrained L\'evy models and produce the respective prices of the exotic products.
This test has the aim of investigating the ability of unconstrained models to reduce the fitting trade-off between marginal and dependence structures.
As anticipated, the calibration of unconstrained models requires a joint procedure as described in Eq.\ \eqref{joint calibration}, because of the increased interconnection between marginal and dependence parameters.
Also, as a result of our first run of calibration, we mostly set $\epsilon=0.1$, and when marginal fit results too poor, we use $\epsilon=0.2$.
Note that joint calibration can hardly set lower maximum correlation errors allowed, such as the $\epsilon=0.01$ of the dependence calibration of constrained BB models (Eq.\ \eqref{BB dependence-calibration}), because it is a high-dimensional problem with computationally expensive objective functions.

\begin{figure} 
\centering
\subfloat[Marginal calibration errors of WP1 underlyings.
]{\includegraphics[scale=0.5]{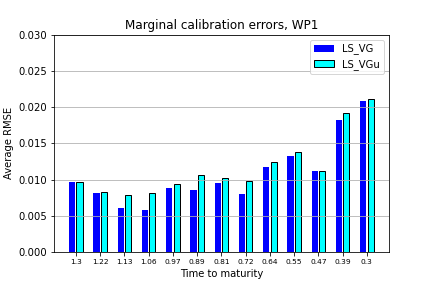}}
\hfill
\subfloat[Dependence calibration errors of WP1 underlyings.
]{\includegraphics[scale=0.5]{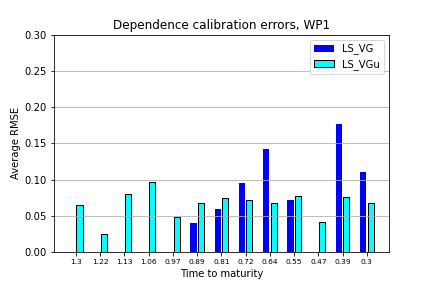}}
\newline
\subfloat[Marginal calibration errors of WP2 underlyings.
]{\includegraphics[scale=0.5]{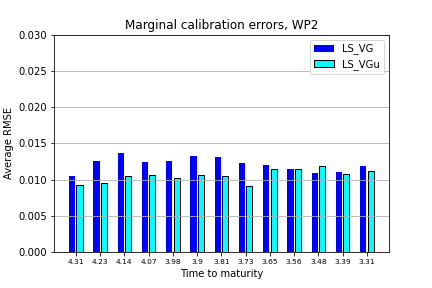}}
\hfill
\subfloat[Dependence calibration errors of WP2 underlyings.
]{\includegraphics[scale=0.5]{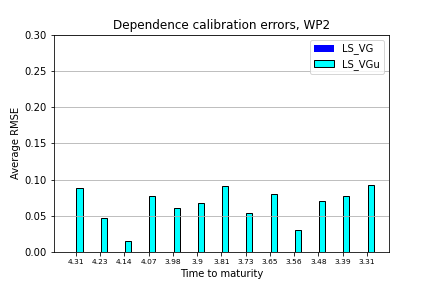}}
\newline
\subfloat[Marginal calibration errors of WP3 underlyings.
]{\includegraphics[scale=0.5]{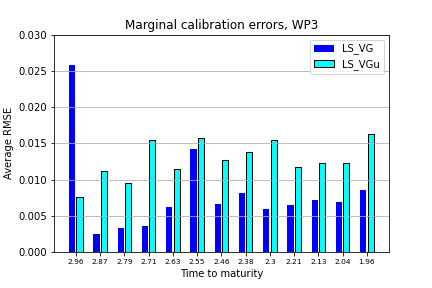}}
\hfill
\subfloat[Dependence calibration errors of WP3 underlyings.
]{\includegraphics[scale=0.5]{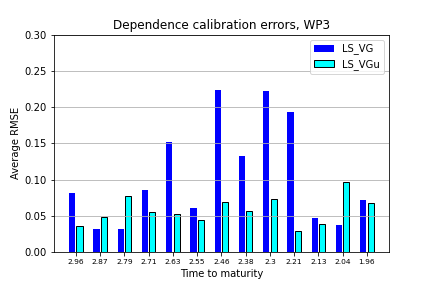}}
\caption{
\small
LS-VG (constrained) and LS-VGu (unconstrained) calibration errors for the 13 considered times to maturity. Left-hand side figures represent average RMSEs between model and market volatilities of the basket of underlyings of each WP. Right-hand side figures represent average RMSEs between model and market correlations of the set of underlying pairs of each WP.
}
\label{fig:VGuLS calibration errors}
\end{figure}

\begin{figure} 
\centering
\subfloat[Marginal calibration errors of WP1 underlyings.
]{\includegraphics[scale=0.5]{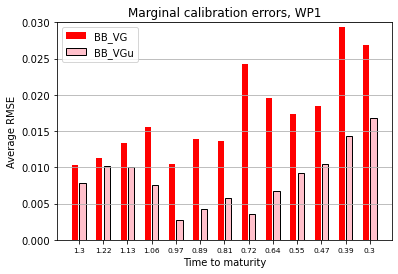}}
\hfill
\subfloat[Dependence calibration errors of WP1 underlyings.
]{\includegraphics[scale=0.5]{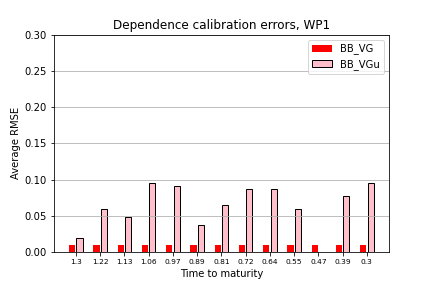}}
\newline
\subfloat[Marginal calibration errors of WP2 underlyings.
]{\includegraphics[scale=0.5]{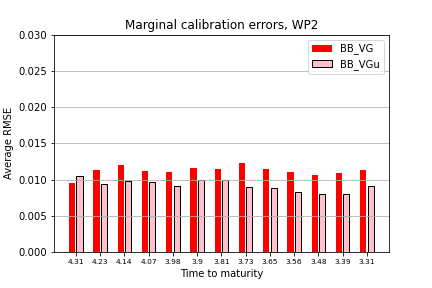}}
\hfill
\subfloat[Dependence calibration errors of WP2 underlyings.
]{\includegraphics[scale=0.5]{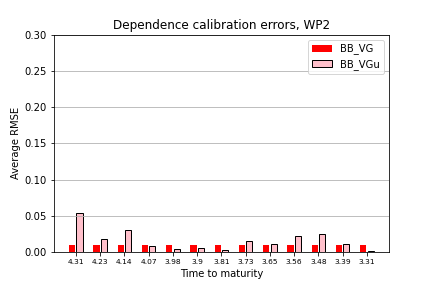}}
\newline
\subfloat[Marginal calibration errors of WP3 underlyings.
]{\includegraphics[scale=0.5]{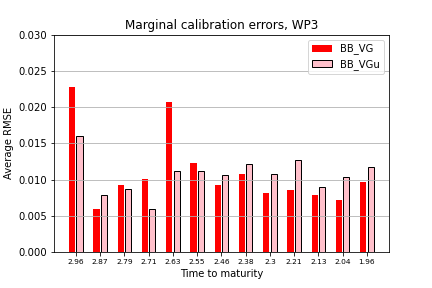}}
\hfill
\subfloat[Dependence calibration errors of WP3 underlyings.
]{\includegraphics[scale=0.5]{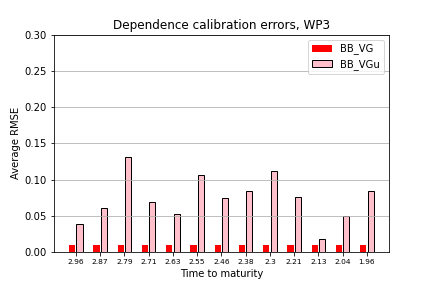}}
\caption{
\small
BB-VG (constrained) and BB-VGu (unconstrained) calibration errors for the 13 considered times to maturity. Left-hand side figures represent average RMSEs between model and market volatilities of the basket of underlyings of each WP. Right-hand side figures represent average RMSEs between model and market correlations of the set of underlying pairs of each WP.
}
\label{fig:VGuBB calibration errors}
\end{figure}

\begin{figure} 
\centering
\subfloat[Marginal calibration errors of WP1 underlyings.
]{\includegraphics[scale=0.5]{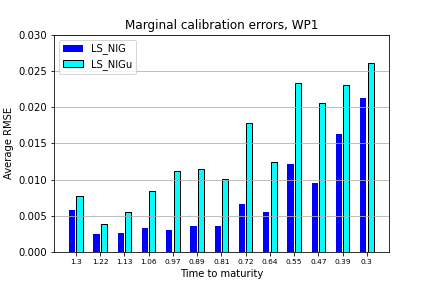}}
\hfill
\subfloat[Dependence calibration errors of WP1 underlyings.
]{\includegraphics[scale=0.5]{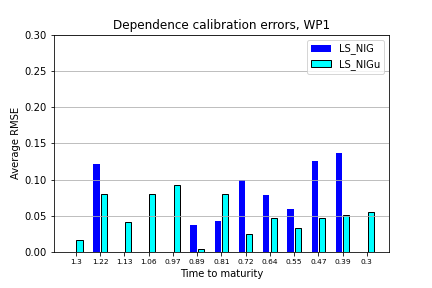}}
\newline
\subfloat[Marginal calibration errors of WP2 underlyings.
]{\includegraphics[scale=0.5]{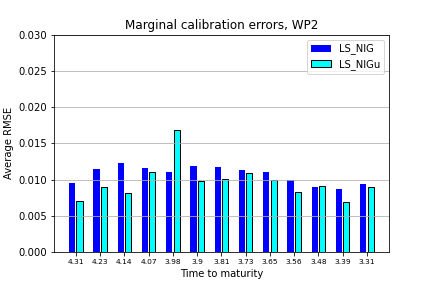}}
\hfill
\subfloat[Dependence calibration errors of WP2 underlyings.
]{\includegraphics[scale=0.5]{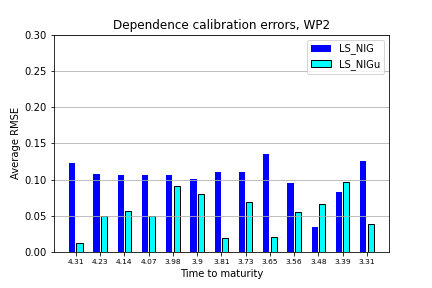}}
\newline
\subfloat[Marginal calibration errors of WP3 underlyings.
]{\includegraphics[scale=0.5]{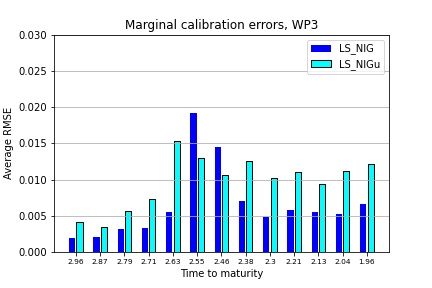}}
\hfill
\subfloat[Dependence calibration errors of WP3 underlyings.
]{\includegraphics[scale=0.5]{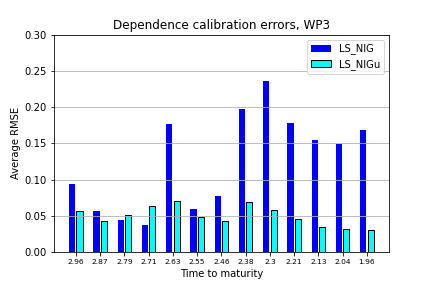}}
\caption{
\small
LS-NIG (constrained) and LS-NIGu (unconstrained) calibration errors for the 13 considered times to maturity. Left-hand side figures represent average RMSEs between model and market volatilities of the basket of underlyings of each WP. Right-hand side figures represent average RMSEs between model and market correlations of the set of underlying pairs of each WP.
}
\label{fig:NIGuLS calibration errors}
\end{figure}

\begin{figure} 
\centering
\subfloat[Marginal calibration errors of WP1 underlyings.
]{\includegraphics[scale=0.5]{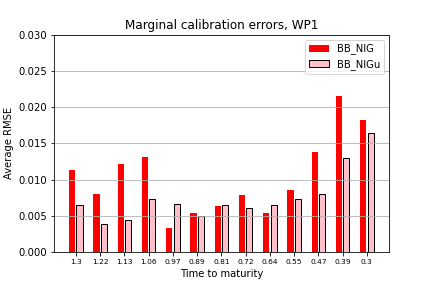}}
\hfill
\subfloat[Dependence calibration errors of WP1 underlyings.
]{\includegraphics[scale=0.5]{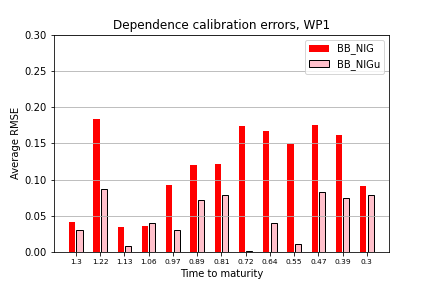}}
\newline
\subfloat[Marginal calibration errors of WP2 underlyings.
]{\includegraphics[scale=0.5]{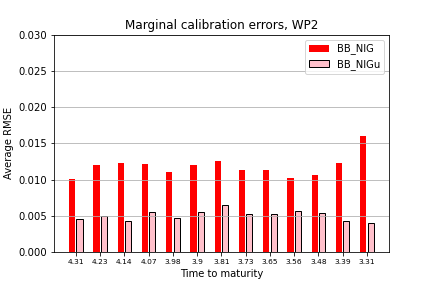}}
\hfill
\subfloat[Dependence calibration errors of WP2 underlyings.
]{\includegraphics[scale=0.5]{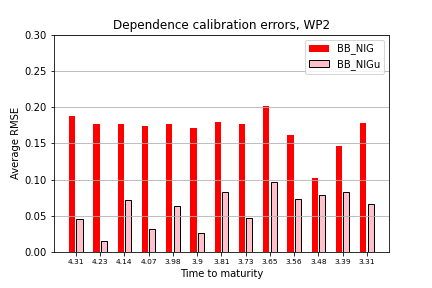}}
\newline
\subfloat[Marginal calibration errors of WP3 underlyings.
]{\includegraphics[scale=0.5]{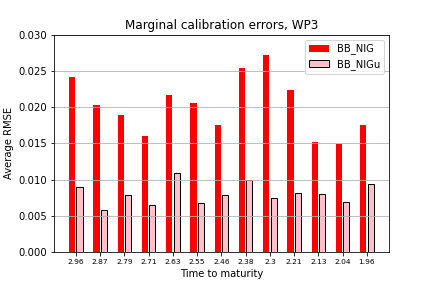}}
\hfill
\subfloat[Dependence calibration errors of WP3 underlyings.
]{\includegraphics[scale=0.5]{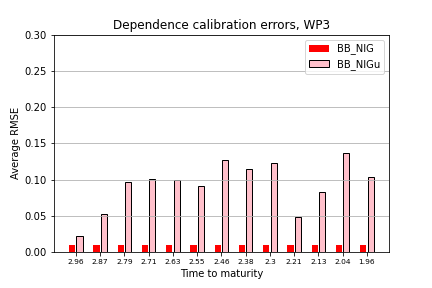}}
\caption{
\small
BB-NIG (constrained) and BB-NIGu (unconstrained) calibration errors for the 13 considered times to maturity. Left-hand side figures represent average RMSEs between model and market volatilities of the basket of underlyings of each WP. Right-hand side figures represent average RMSEs between model and market correlations of the set of underlying pairs of each WP.
}
\label{fig:NIGuBB calibration errors}
\end{figure}

\begin{figure} 
\centering
\subfloat[WP1 price percentage differences of constrained models.
]{\includegraphics[scale=0.5]{PercDiff_XS1945967153_Rho-inf.png}}
\hfill
\subfloat[WP1 price percentage differences of unconstrained models.
]{\includegraphics[scale=0.5]{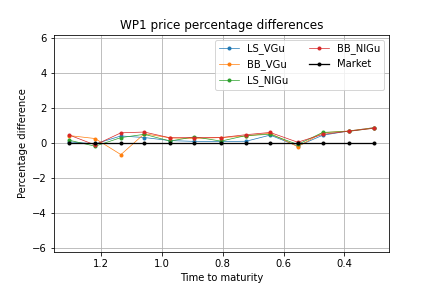}}
\newline
\subfloat[WP2 price percentage differences of constrained models.
]{\includegraphics[scale=0.5]{PercDiff_XS2115185295_Rho-inf.png}}
\hfill
\subfloat[WP2 price percentage differences of unconstrained models.
]{\includegraphics[scale=0.5]{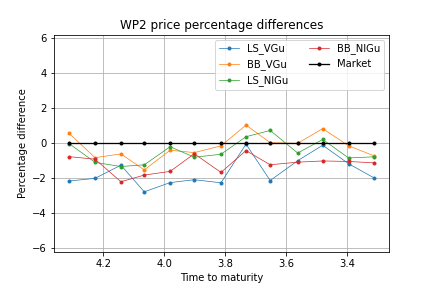}}
\newline
\subfloat[WP3 price percentage differences of constrained models.
]{\includegraphics[scale=0.5]{PercDiff_XS2402145127_Rho-inf.png}}
\hfill
\subfloat[WP3 price percentage differences of unconstrained models.
]{\includegraphics[scale=0.5]{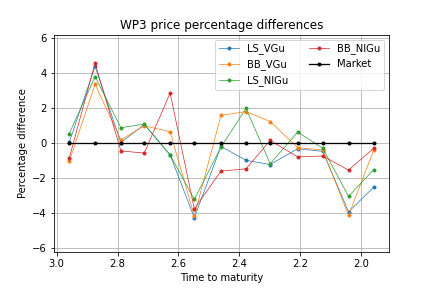}}
\caption{
\small
Percentage differences between model and market prices, on the 13 considered times to maturity. Percentage differences are computed as (Price $-$ Market Price) / Market Price $\times$ 100.
}
\label{fig:prices constrained vs unconstrained}
\end{figure}

In order to provide a visible comparison between constrained and unconstrained models, each calibration plot displays fitting errors of a constrained L\'evy model and its unconstrained version, as shown in Figs.\ \ref{fig:VGuLS calibration errors}, \ref{fig:VGuBB calibration errors}, \ref{fig:NIGuLS calibration errors} and \ref{fig:NIGuBB calibration errors}. 
As can be seen, in most cases the two approaches produce comparable performances, which is also evidenced by the pricing results of Fig.\ \ref{fig:prices constrained vs unconstrained}.
However, a notable exception is given by BB-NIG, whose calibration plots (Fig.\ \ref{fig:NIGuBB calibration errors}) show a better fit of its unconstrained version.
In essence, the goal of unconstrained models to reduce the marginal vs dependence trade-off on LS models is not fully reached.
However, the results of BB models reward the unconstrained structure, at least from the point of view of pure model fitting.
As a matter of fact, not only the authors of BB processes started preferring the unconstrained versions of their model in recent works (as \citealp{ballotta2017multivariate} and \citealp{ballotta2019estimation}), but the convenience of choosing such construction also has theoretical explanations.
In fact, constrained BB embeds exigent restrictions (see Eqs.\ \eqref{VG-conv-restrictions-1} and \eqref{NIG-conv-restrictions-1}) to preserve known marginal laws, as opposed to the simpler bounds on parameter $a$ for LS models (see Eqs.\ \eqref{Gamma-Sub} and \eqref{IG-Sub}).
Overall, removing constraints in BB models is a more valid choice than it is for LS models.

\subsection{Key Implications of the Empirical Results}\label{Implications}
The results reported in the empirical analysis allow us to draw some implications that can be relevant for any application of multivariate L\'evy models. 

First, we highlight the main strengths shared by all the analyzed processes. 
A core feature of these constructions is given by the explicit expressions of the characteristic functions, which allow to recover semi-closed formulas for plain-vanilla options, or more broadly, for the terminal density function.
In our analysis, such semi-closed formulas correspond to calibrations that are fast enough to be employed in real financial contexts.
Also, the factor-based construction via subordination of Brownian motions permits a simple and sound simulation scheme \lq by components\rq.
That is, we sample from the laws of well-known univariate subordinators, for which efficient simulation algorithms are available, and separately, sample from the Brownian motions.
Moreover, Fig.\ \ref{fig:prices constrained vs unconstrained} shows that all the L\'evy constructions accurately capture the tails of the short-term distributions of log-returns.
As such, it is a good choice to employ jump processes to model short-term extreme events in general, even at the multivariate level. 

Second, following the discussion of Subsection \ref{Calibration Choices} it is advisable to perform a preliminary calibration run on the dataset of interest.
This is particularly important to select the domains of the model parameters within reasonable ranges, speeding up the convergence of calibrations that can be expensive in high dimensions.
Moreover, it is crucial to identify the time periods in which the distribution of the stochastic factors is most important.
As the payoff of the analyzed exotic options is concentrated at maturity, we have selected -- supported by the preliminary pricing run -- the calibration weights in order to focus the estimation of the model on the terminal distribution.
If, alternatively, the model is calibrated to the realized measure for risk management purposes, it can be desirable to focus the fit on a particular historical period.
While L\'evy models are exceptionally effective in representing a single-maturity law, their stationary nature requires a wise choice of the time period of interest.

Further considerations relate to the marginal constraints of the models.
Setting convolution conditions keeps parameters highly interpretable, which can be convenient to better understand the model and to meet possible requirements of regulators.
However, the results of Subsection \ref{Constrained vs Unconstrained Margins} have demonstrated that for some L\'evy models it is desirable to remove marginal constraints.
As a general rule, satisfying a few inequality constraints (see, e.g., Eq.\ \eqref{IG-Sub}) comes at almost no cost, while in presence of equality constraints (see Eq.\ \eqref{NIG-conv-restrictions-1}) imposing marginals of known class risks to excessively reduce model fit. 
If calibration shows a large trade-off between the fit of marginal and dependence parameters, the user can either remove restrictions, or keep convolution conditions and employ alternative calibration methods (see Subsection \ref{Model-Specific Calibrations}) to regulate the amount of error to place between margins and dependence.
We believe these choices mainly stand on the problem-specific desire of interpretability, and the relative importance between the marginal and the dependence fit (in our case, determined by the exotic payoffs).

\section{Conclusions}\label{Conclusions}

In this paper we investigate several multivariate L\'evy models following the constructions of \cite{luciano2010multivariate} and \cite{ballotta2016multivariate}, and we study calibration methods to fit them.
We perform an extensive empirical analysis to test the ability of the models to fit market data, give a fair valuation to exotic contracts, and suitably capture
dependence features.
Our results suggest that constrained LS models are a more stable choice than constrained BB models, as BB construction is sensitive to model-specific convolution conditions.
Also, the ability of LS processes to embed nonlinear dependence makes them richer constructions, better suited to fit the joint dynamics of log-returns.
On the other hand, the unconstrained BB models prove to be able to reduce the trade-off between marginal and correlation fit observed in constrained models.
Still, while unconstrained processes retain a flexible structure, they also have less interpretable parameters and do not allow for two-step calibrations.
As a consequence, choosing to remove marginal restrictions should depend on the priorities of the user, with regards to model fitting, computational effort and interpretability.

As prices computed by L\'evy models tend to lose accuracy on path-dependent products with medium-long maturity, future research could focus on additive non-stationary processes.
For example, it would be relevant to assess and compare versions of LS and BB models that embed Sato margins (studied by \citealp{marena2018multivariate} and \citealp{boen2019building}).
However, as these processes keep the same dependence structure of the stationary versions, a further direction for research can be to test models that allow for more realistic time-dependent correlations (see, e.g., \citealp{semeraro2022multivariate}, \citealp{AMICI20251004}).
We finally highlight a relatively recent area of study focused on time-changed L\'evy processes (see, e.g., \citealp{ballotta2022smiles}, and at the multivariate level, \citealp{fontana2022cbi}), which are valid competitors of Sato-like processes in exotic option pricing.
\section*{Compliance with Ethical Standards}
Authors declare that they have no conflict of interest.

\section*{Acknowledgements}
Patrizia Semeraro is partially supported by GNAMPA 2023 project: \lq\lq Limiting Behavior of Stochastic Dynamics in the Schelling Segregation Model\rq\rq\: CUP\_E53C22001930001.

\clearpage

\begin{thebibliography}{}

\bibitem[Alfeus et~al., 2020]{alfeus2020regularization}
Alfeus, M., He, X.-J., and Zhu, S.-P. (2020).
\newblock Regularization effect on model calibration.
\newblock {\em Journal of Risk}, 24(3).

\bibitem[Amici et~al., 2025]{AMICI20251004}
Amici, G., Ballotta, L., and Semeraro, P. (2025).
\newblock Multivariate additive subordination with applications in finance.
\newblock {\em European Journal of Operational Research}, 321(3):1004--1020.

\bibitem[Ballotta and Bonfiglioli, 2016]{ballotta2016multivariate}
Ballotta, L. and Bonfiglioli, E. (2016).
\newblock Multivariate asset models using {L\'e}vy processes and applications.
\newblock {\em The European Journal of Finance}, 22(13):1320--1350.

\bibitem[Ballotta et~al., 2017]{ballotta2017multivariate}
Ballotta, L., Deelstra, G., and Ray{\'e}e, G. (2017).
\newblock Multivariate {FX} models with jumps: Triangles, quantos and implied
  correlation.
\newblock {\em European Journal of Operational Research}, 260(3):1181--1199.

\bibitem[Ballotta et~al., 2019]{ballotta2019estimation}
Ballotta, L., Fusai, G., Loregian, A., and Perez, M.~F. (2019).
\newblock Estimation of multivariate asset models with jumps.
\newblock {\em Journal of Financial and Quantitative Analysis},
  54(5):2053--2083.

\bibitem[Ballotta and Ray{\'e}e, 2022]{ballotta2022smiles}
Ballotta, L. and Ray{\'e}e, G. (2022).
\newblock Smiles \& smirks: Volatility and leverage by jumps.
\newblock {\em European Journal of Operational Research}, 298(3):1145--1161.

\bibitem[Barndorff-Nielsen, 1997]{barndorff1997processes}
Barndorff-Nielsen, O.~E. (1997).
\newblock Processes of normal inverse Gaussian type.
\newblock {\em Finance and stochastics}, 2:41--68.

\bibitem[Barndorff-Nielsen et~al., 2001]{barndorff2001multivariate}
Barndorff-Nielsen, O.~E., Pedersen, J., and Sato, K. (2001).
\newblock Multivariate subordination, self-decomposability and stability.
\newblock {\em Advances in Applied Probability}, 33(1):160--187.

\bibitem[Bertsimas et~al., 2001]{bertsimas2001hedging}
Bertsimas, D., Kogan, L., and Lo, A.~W. (2001).
\newblock Hedging derivative securities and incomplete markets: An
  $\epsilon$-arbitrage approach.
\newblock {\em Operations research}, 49(3):372--397.

\bibitem[Bianchi et~al., 2022]{bianchi2022welcome}
Bianchi, M.~L., Hitaj, A., and Tassinari, G.~L. (2022).
\newblock A welcome to the jungle of continuous-time multivariate
  non-{G}aussian models based on {L\'e}vy processes applied to finance.
\newblock {\em Annals of Operations Research}, pages 1--42.

\bibitem[Blomvall and Hagenbj{\"o}rk, 2022]{blomvall2022reducing}
Blomvall, J. and Hagenbj{\"o}rk, J. (2022).
\newblock Reducing transaction costs for interest rate risk hedging with
  stochastic programming.
\newblock {\em European Journal of Operational Research}, 302(3):1282--1293.

\bibitem[Boen and Guillaume, 2019]{boen2019building}
Boen, L. and Guillaume, F. (2019).
\newblock Building multivariate {S}ato models with linear dependence.
\newblock {\em Quantitative Finance}, 19(4):619--645.

\bibitem[Boyarchenko and Levendorskii, 2002]{boyarchenko2002perpetual}
Boyarchenko, S.~I. and Levendorskii, S.~Z. (2002).
\newblock Perpetual {A}merican options under {L\'e}vy processes.
\newblock {\em SIAM Journal on Control and Optimization}, 40(6):1663--1696.

\bibitem[Broadie and Kaya, 2006]{broadie2006exact}
Broadie, M. and Kaya, {\"O}. (2006).
\newblock Exact simulation of stochastic volatility and other affine jump
  diffusion processes.
\newblock {\em Operations research}, 54(2):217--231.

\bibitem[Camc$\iota$ and {\c{C}}.~P$\iota$nar, 2009]{camciota2009pricing}
Camc$\iota$, A. and {\c{C}}.~P$\iota$nar, M. (2009).
\newblock Pricing {A}merican contingent claims by stochastic linear
  programming.
\newblock {\em Optimization}, 58(6):627--640.

\bibitem[Chiarolla et~al., 2015]{chiarolla2015optimal}
Chiarolla, M.~B., Ferrari, G., and Stabile, G. (2015).
\newblock Optimal dynamic procurement policies for a storable commodity with
  {L\'e}vy prices and convex holding costs.
\newblock {\em European Journal of Operational Research}, 247(3):847--858.

\bibitem[Cont and Ben~Hamida, 2004]{cont2004recovering}
Cont, R. and Ben~Hamida, S. (2004).
\newblock Recovering volatility from option prices by evolutionary
  optimization.
\newblock {\em The Journal of Computational Finance}, 8(4):43--76.

\bibitem[Cont and Tankov, 2004]{cont2004nonparametric}
Cont, R. and Tankov, P. (2004).
\newblock Nonparametric calibration of jump-diffusion option pricing models.
\newblock {\em The Journal of Computational Finance}, 7:1--49.

\bibitem[Cont and Tankov, 2006]{cont2006retrieving}
Cont, R. and Tankov, P. (2006).
\newblock Retrieving {L\'e}vy processes from option prices: Regularization of
  an ill-posed inverse problem.
\newblock {\em SIAM Journal on Control and Optimization}, 45(1):1--25.

\bibitem[Cr{\'e}pey, 2010]{crepey2010tikhonov}
Cr{\'e}pey, S. (2010).
\newblock Tikhonov regularization.
\newblock {\em Encyclopedia of Quantitative Finance}, pages 1807--1812.

\bibitem[Dai et~al., 2016]{dai2016calibration}
Dai, M., Tang, L., and Yue, X. (2016).
\newblock Calibration of stochastic volatility models: a {T}ikhonov
  regularization approach.
\newblock {\em Journal of Economic Dynamics and Control}, 64:66--81.

\bibitem[Egger and Engl, 2005]{egger2005tikhonov}
Egger, H. and Engl, H.~W. (2005).
\newblock Tikhonov regularization applied to the inverse problem of option
  pricing: convergence analysis and rates.
\newblock {\em Inverse Problems}, 21(3):1027.

\bibitem[Fang and Oosterlee, 2009]{fang2009novel}
Fang, F. and Oosterlee, C.~W. (2009).
\newblock A novel pricing method for european options based on fourier-cosine
  series expansions.
\newblock {\em SIAM Journal on Scientific Computing}, 31(2):826--848.

\bibitem[Fontana et~al., 2022]{fontana2022cbi}
Fontana, C., Gnoatto, A., and Szulda, G. (2022).
\newblock {CBI}-time-changed {L\'e}vy processes for multi-currency modeling.
\newblock {\em Annals of Operations Research}, pages 1--26.

\bibitem[Fu and Yang, 2012]{fu2012equilibruim}
Fu, J. and Yang, H. (2012).
\newblock Equilibruim approach of asset pricing under {L\'e}vy process.
\newblock {\em European Journal of Operational Research}, 223(3):701--708.

\bibitem[Glasserman and Liu, 2010]{glasserman2010sensitivity}
Glasserman, P. and Liu, Z. (2010).
\newblock Sensitivity estimates from characteristic functions.
\newblock {\em Operations Research}, 58(6):1611--1623.

\bibitem[Godin, 2016]{godin2016minimizing}
Godin, F. (2016).
\newblock Minimizing {CVaR} in global dynamic hedging with transaction costs.
\newblock {\em Quantitative Finance}, 16(3):461--475.

\bibitem[Guillaume, 2013]{guillaume2013alpha}
Guillaume, F. (2013).
\newblock The $\alpha${VG} model for multivariate asset pricing: Calibration
  and extension.
\newblock {\em Review of Derivatives Research}, 16:25--52.

\bibitem[K{\"u}chler and Tappe, 2013]{kuchler2013tempered}
K{\"u}chler, U. and Tappe, S. (2013).
\newblock Tempered stable distributions and processes.
\newblock {\em Stochastic Processes and their Applications},
  123(12):4256--4293.

\bibitem[Li and Linetsky, 2013]{li2013optimal}
Li, L. and Linetsky, V. (2013).
\newblock Optimal stopping and early exercise: an eigenfunction expansion
  approach.
\newblock {\em Operations Research}, 61(3):625--643.

\bibitem[Luciano and Semeraro, 2010]{luciano2010multivariate}
Luciano, E. and Semeraro, P. (2010).
\newblock Multivariate time changes for {L\'e}vy asset models: Characterization
  and calibration.
\newblock {\em Journal of Computational and Applied Mathematics},
  233(8):1937--1953.

\bibitem[Madan et~al., 1998]{madan1998variance}
Madan, D.~B., Carr, P.~P., and Chang, E.~C. (1998).
\newblock The variance gamma process and option pricing.
\newblock {\em Review of Finance}, 2(1):79--105.

\bibitem[Marena et~al., 2018a]{marena2018multivariate}
Marena, M., Romeo, A., and Semeraro, P. (2018a).
\newblock Multivariate factor-based processes with {S}ato margins.
\newblock {\em International Journal of Theoretical and Applied Finance},
  21(01):1850005.

\bibitem[Marena et~al., 2018b]{marena2018pricing}
Marena, M., Romeo, A., and Semeraro, P. (2018b).
\newblock Pricing multivariate barrier reverse convertibles with factor-based
  subordinators.
\newblock {\em Journal of Computational Finance}, 21(5).

\bibitem[Ruf and Wang, 2020]{ruf2020neural}
Ruf, J. and Wang, W. (2020).
\newblock Neural networks for option pricing and hedging: a literature review.
\newblock {\em Journal of Computational Finance}.

\bibitem[Sato, 1999]{ken1999levy}
Sato, K. (1999).
\newblock {\em L{\'e}vy processes and infinitely divisible distributions}.
\newblock Cambridge University Press.

\bibitem[Semeraro, 2008]{semeraro2008multivariate}
Semeraro, P. (2008).
\newblock A multivariate variance gamma model for financial applications.
\newblock {\em International Journal of Theoretical and Applied Finance},
  11(01):1--18.

\bibitem[Semeraro, 2022]{semeraro2022multivariate}
Semeraro, P. (2022).
\newblock Multivariate tempered stable additive subordination for financial
  models.
\newblock {\em Mathematics and Financial Economics}, 16(4):685--712.

\bibitem[Storn and Price, 1997]{storn1997differential}
Storn, R. and Price, K. (1997).
\newblock Differential evolution: a simple and efficient heuristic for global
  optimization over continuous spaces.
\newblock {\em Journal of Global Optimization}, 11(4):341.

\bibitem[Tang et~al., 2023]{tang2023structural}
Tang, Z., Zhong, B., Zhou, L., and Shen, C. (2023).
\newblock Structural credit risk model driven by {L\'e}vy process under knight
  uncertainty.
\newblock {\em Annals of Operations Research}, 326(1):281--294.

\bibitem[Villaverde, 2004]{villaverde2004hedging}
Villaverde, M. (2004).
\newblock Hedging european and barrier options using stochastic optimization.
\newblock {\em Quantitative Finance}, 4(5):549--557.

\end{thebibliography}

\clearpage
\appendix

\section{
Details of the model constructions
}\label{app:Models}
\subsection{Constrained LS-Variance Gamma}\label{LS-VG}

As it can be seen from the LS process construction of Subsection \ref{LS Models}, the L\'evy class of $Y_j(t)$ is given by the specifications of subordinators. Here we illustrate the case where each asset log-return follows a Variance Gamma process \citep{madan1998variance}.
Let $X_1(t), ..., X_n(t), Z(t)$ be Gamma processes such that
\begin{equation*}
    X_j \sim \Gamma \left(\frac{1}{\kappa_j} - a, \frac{1}{\kappa_j} \right),
    \hspace{15pt}
    Z \sim \Gamma \left(a, 1 \right),
    \hspace{15pt}
    \text{with}
    \hspace{15pt}
    0 < a < \min_j \left( \frac{1}{\kappa_j} \right),
    \hspace{5pt}
    \kappa_j > 0.
\end{equation*}
Then, by the closure property of convolution of Gamma distributions, and following Eq.\ \eqref{G_j}, we get 
$$
G_j \sim \Gamma \left(\frac{1}{\kappa_j}, \frac{1}{\kappa_j} \right).
$$
\cite{luciano2010multivariate} proved that the resulting one-dimensional margin $Y_j(t)$ follows a VG$(\mu_j, \sigma_j, \kappa_j)$, i.e.,
\begin{equation}\label{VG-CF-copy}
    \phi_{Y_j} (u) = \left( 1 - i u \mu_j \kappa_j + \frac{1}{2} u^2 \sigma^2_j \kappa_j \right)
    ^{-\kappa^{-1}_j},
    \hspace{15pt}
    u \in \mathbb{R}
    ,
\end{equation}
while the multivariate process $\boldsymbol{Y} (t)$ is an LS-VG process with parameters \\$( \boldsymbol{\mu}, \boldsymbol{\sigma}, \boldsymbol{\kappa}, a, \{ \rho_{ij} \}_{i \neq j} )$ and characteristic function
\begin{equation}\label{LS-VG-CF}
\phi _{\boldsymbol{Y}}(\boldsymbol{u})=
\prod_{j=1}^{n}\left[ 1-{\kappa_{j}\left(i\mu _{j}u_{j}-\frac{1}{2}\sigma _{j}^{2}u_{j}^{2}\right)}\right]
^{-\left( \kappa^{-1}_j-a\right) }
\left[ 1-\left( i\boldsymbol{u}^{T} \boldsymbol{\mu }^{\rho }-\frac{1}{2}\boldsymbol{u}^{T}\boldsymbol{\Sigma }^{\rho }\boldsymbol{u}\right) \right] ^{-a}
\hspace{-8pt}
,
\hspace{1pt}
\boldsymbol{u} \in \mathbb{R}^n.
\end{equation}%
The number of parameters of the time-1 distribution is $1 + 3n + \frac{n(n-1)}{2}$ and the correlation, for each pair $(i,j)$ of assets, is given by
\begin{equation*}
    \rho_{\boldsymbol{Y}} (i,j)
    =
    \frac{ a\left(
    \mu_i \mu_j \kappa_i \kappa_j
    +
    \rho_{ij} \sigma_i \sigma_j \sqrt{\kappa_i} \sqrt{\kappa_j} \right)
    }{
    \sqrt{ (\sigma^2_i + \mu^2_i \kappa_i)
    (\sigma^2_j + \mu^2_j \kappa_j) }
    }.
\end{equation*}
Bounds on the correlation coefficient are thoroughly discussed by \cite{marena2018pricing}.
They observed that the maximum achievable level of correlation is linked to the marginal with the highest $\kappa$ parameter. As $\kappa$ drives the kurtosis, there exists a trade-off between the kurtosis marginal fit and the correlation admissible range.
However, an interesting feature is that dependence structure allows for non-linear dependence, which can be easily observed in case of symmetric marginals (i.e., $\mu_i= \mu_j = 0$) and uncorrelated Brownian motions. In this circumstance, we would get the correlation coefficient equal to 0 but still have non-zero dependence regulated by parameter $a$.

\subsection{Unconstrained LS-Variance Gamma}\label{LS-VG-uncLS}

As can be seen from \eqref{Gamma-Sub}, 
parameter $a$ is bounded above from a term related to $\kappa_j$, that is a driver of marginal kurtosis.
The constraint on $a$ comes from the convolution condition needed to exclude $a$ from the marginal parametrization and to get an unbiased gamma subordinator, as in the traditional VG process.
As $a$ drives part of correlation, this interconnection among parameters is a source of trade-off between marginal and joint fit.
To deal with this issue, the convolution restriction can be relaxed, at the expense that one-dimensional margins do not have the traditional VG parametrization.
To this aim, we can introduce a new parameter vector $\boldsymbol{\alpha} = (\alpha_1, ..., \alpha_n)$, that provides more modeling flexibility, as follows.
Let $X_1(t), ..., X_n(t), Z(t)$ be Gamma processes such that
\begin{equation}\label{Gamma-Sub-uncLS}
    X_j \sim \Gamma \left(\alpha_j , \frac{1}{\kappa_j} \right),
    \hspace{15pt}
    Z \sim \Gamma \left(a, 1 \right),
    \hspace{15pt}
    \text{with}
    \hspace{15pt}
    a, \alpha_j, \kappa_j > 0.
\end{equation}
Then, by the closure property of convolution of Gamma distributions, and following Eq.\ \eqref{G_j}, we get 
$$
G_j \sim \Gamma \left(\alpha_j + a, \frac{1}{\kappa_j} \right).
$$
As a consequence, the characteristic function of the marginal distribution is
\begin{equation*}
    \phi_{Y_j} (u) = \left( 1 - i u \mu_j \kappa_j + \frac{1}{2} u^2 \sigma^2_j \kappa_j \right)
    ^{-(\alpha_j+a)},
    \hspace{15pt}
    u \in \mathbb{R}
    .
\end{equation*}
The characteristic function of the resulting multivariate process $\boldsymbol{Y}(t)$ follows from \eqref{LS-CF}, and the correlation, for each pair $(i,j)$ of assets, is given by
\begin{equation*}
    \rho_{\boldsymbol{Y}} (i,j)
    =
    \frac{ a\left(
    \mu_i \mu_j \kappa_i \kappa_j
    +
    \rho_{ij} \sigma_i \sigma_j \sqrt{\kappa_i} \sqrt{\kappa_j} \right)
    }{
    \sqrt{ \kappa_i (\alpha_i + a) (\sigma^2_i + \mu^2_i \kappa_i) \:
    \kappa_j (\alpha_j + a)(\sigma^2_j + \mu^2_j \kappa_j) }
    }.
\end{equation*}
Note that here $a$ does not impact on the correlation range as it does in the constrained LS-VG model, since it is part of the marginal parametrization and now also appears in the denominator of $\rho_{\boldsymbol{Y}} (i,j)$. Therefore, having removed the original upper bound on $a$ does not, per se, increase the correlation admissible range. However, we are still interested in checking whether the interaction between these unconstrained parameters can, in practice, produce a more flexible dependence structure.

\subsection{Constrained LS-Normal Inverse Gaussian}\label{LS-NIG}

Another well-known type of L\'evy process is given by Normal Inverse Gaussian process \citep{barndorff1997processes}. Here we show how an LS process can be built such that its margins are NIG processes. Let $X_1(t), ..., X_n(t), Z(t)$ be Inverse Gaussian processes,
\begin{equation*}
    X_j \sim \text{IG} \left(1 - a \sqrt{\kappa_j}, \frac{1}{\sqrt{\kappa_j}} \right),
    \hspace{15pt}
    Z \sim \text{IG} \left(a, 1 \right),
    \hspace{15pt}
    \text{with}
    \hspace{15pt}
    0 < a < \min_j \left( \frac{1}{\sqrt{\kappa}_j} \right),
    \hspace{5pt}
    \kappa_j > 0.
\end{equation*}
Then, by the closure property of convolution of Inverse Gaussian distributions, and following Eq.\ \eqref{G_j}, we get 
$$
G_j \sim \text{IG} \left(1, \frac{1}{\sqrt{\kappa_j}} \right).
$$
\cite{luciano2010multivariate} proved that the resulting one-dimensional margin $Y_j(t)$ follows a NIG$(\beta_j, \delta_j, \gamma_j)$, with $-\gamma_j<\beta_j<\gamma_j$, $\delta_j>0$, $\gamma_j>0$, i.e.,
\begin{equation}\label{NIG-CF-copy}
    \phi_{Y_j} (u) = \exp \left\{ -\delta_j \left(\sqrt{\gamma^2_j - (\beta_j + i u)^2} - \sqrt{\gamma^2_j - \beta^2_j}\right) \right\}
    ,
    \hspace{15pt}
    u \in \mathbb{R}
    ,
\end{equation}
where, applying relations
$$
\mu_j = \beta_j \delta^2,
\hspace{3pt}
\sigma_j = \delta_j,
\hspace{3pt}
\kappa_j = [\delta_j^2 (\gamma_j^2 - \beta_j^2)]^{-1},
$$
we get the same expression of Eq.\ \eqref{Y_j}.
$\boldsymbol{Y} (t)$ is then a LS-NIG$( \boldsymbol{\beta}, \boldsymbol{\delta}, \boldsymbol{\gamma}, a, \{ \rho_{ij} \}_{i \neq j} )$ with $1 + 3n + \frac{n(n-1)}{2}$ parameters, time-1 characteristic function 
\begin{equation*}
\begin{split}
\phi _{\boldsymbol{Y}}(\boldsymbol{u})=
& \exp \left\{ -\sum_{j=1}^{n} \left(1-a\sqrt{\kappa_{j}} \right)\left( \sqrt{-2 \left(i\beta _{j}\delta _{j}^{2}u_{j}-\frac{1}{2}\delta _{j}^{2}u_{j}^{2} \right)+\frac{1}{\kappa_{j}}} - \frac{1}{\sqrt{\kappa_{j}}} \right) \right\} \cdot
\\
& \cdot \exp \left\{ -a\left( \sqrt{-2 \left(i\boldsymbol{u}^{T}\boldsymbol{\mu}^{\rho}-\frac{1}{2}\boldsymbol{u}^{T}\boldsymbol{\Sigma}^{\rho}\boldsymbol{u}\right)+1}-1\right) \right\},
\hspace{10pt}
\boldsymbol{u} \in \mathbb{R}^n,
\end{split}%
\end{equation*}
and correlation coefficient
\begin{equation*}
    \rho_{\boldsymbol{Y}} (i,j)
    =
    \frac{ a\left(
    \beta_i \delta_i^2 \kappa_i
    \beta_j \delta_j^2 \kappa_j
    +
    \rho_{ij} \delta_i \sqrt{\kappa_i} \delta_j \sqrt{\kappa_j} \right)
    }{
    \sqrt{
    \gamma_i^2 \delta_i (\gamma_i^2 - \beta_i^2)^{-3/2}
    \:
    \gamma_j^2 \delta_j (\gamma_j^2 - \beta_j^2)^{-3/2}
    }
    }.
\end{equation*}
As far as the  dependence structure is concerned, considerations
similar to those in Subsection \ref{LS-VG} can be made.

\subsection{Unconstrained LS-Normal Inverse Gaussian}\label{LS-NIG-uncLS}

As in the VG case, it is possible to weaken the constraints on $a$ to try to reduce the trade-off between marginal and dependence fit of the NIG specification.
Let $X_1(t), \ldots, X_n(t)$, $Z(t)$ be Inverse Gaussian processes such that
\begin{equation}\label{IG-Sub-uncLS}
    X_j \sim \text{IG} \left(\alpha_j, \frac{1}{\sqrt{\kappa_j}} \right),
    \hspace{15pt}
    Z \sim \text{IG} \left(a, 1 \right),
    \hspace{15pt}
    \text{with}
    \hspace{15pt}
    a, \alpha_j, \kappa_j > 0.
\end{equation}
Then, by the closure property of convolution of Inverse Gaussian distributions, and following Eq.\ \eqref{G_j}, we get 
$$
G_j \sim \text{IG} \left(\alpha_j + a \sqrt{k_j}, \frac{1}{\sqrt{\kappa_j}} \right).
$$
As a consequence, the characteristic function of the marginal distribution is
\begin{equation*}
    \phi_{Y_j} (u) = \exp \left\{ -\delta_j \left(\sqrt{\gamma^2_j - (\beta_j + i u)^2} - \sqrt{\gamma^2_j - \beta^2_j}\right) \left( \alpha_j + a \sqrt{\kappa_j} \right) \right\}
    ,
    \hspace{15pt}
    u \in \mathbb{R}
    ,
\end{equation*}
where $\kappa_j = [\delta_j^2 (\gamma_j^2 - \beta_j^2)]^{-1}$.
The characteristic function of the resulting multivariate process $\boldsymbol{Y}(t)$ follows from \eqref{LS-CF}, and the correlation coefficient is
\begin{equation*}
    \rho_{\boldsymbol{Y}} (i,j)
    =
    \frac{ a\left(
    \beta_i \delta_i^2 \kappa_i
    \beta_j \delta_j^2 \kappa_j
    +
    \rho_{ij} \delta_i \sqrt{\kappa_i} \delta_j \sqrt{\kappa_j} \right)
    }{
    \sqrt{
    (\alpha_i + a \sqrt{\kappa_i})
    \gamma_i^2 \delta_i (\gamma_i^2 - \beta_i^2)^{-3/2}
    \:
    (\alpha_j + a \sqrt{\kappa_j})
    \gamma_j^2 \delta_j (\gamma_j^2 - \beta_j^2)^{-3/2}
    }
    }.
\end{equation*}

\subsection{Constrained BB-Variance Gamma}\label{BB-VG}

The construction of a BB model of VG type is made by assuming that each component follows a VG process. Getting the resulting process with VG margins requires some convolution conditions that follow from  Eq.\ \eqref{conv-conditions}, and that have been found by Ballotta and Bonfiglioli. In particular, let
$
Y_j \sim \text{VG} (\mu_j, \sigma_j, \kappa_j)
,
\hspace{3pt}
X_j \sim \text{VG} (\mu_{X_j}, \sigma_{X_j}, \kappa_{X_j})
,
\hspace{3pt}
Z \sim \text{VG} (\mu_Z, \sigma_Z, \kappa_Z)
$,
with characteristic functions as in Eq.\ \eqref{VG-CF-copy}. Then, to satisfy Eq.\ \eqref{conv-conditions}, the following constraints must hold:
\begin{equation}\label{VG-conv-restrictions-1-copy}
\begin{aligned}
    \kappa_j \mu_j = \kappa_Z b_j \mu_Z
    ,
    \hspace{15pt}
    j = 1, ..., n,
    \\
    \kappa_j \sigma^2_j = \kappa_Z b^2_j \sigma^2_Z
    ,
    \hspace{15pt}
    j = 1, ..., n,
\end{aligned}
\end{equation}
from which we obtain relations
\begin{equation}\label{VG-conv-restrictions-2-copy}
    \mu_j = \mu_{X_j} + b_j \mu_Z
    ,
    \hspace{15pt}
    \sigma_j = \sqrt{ \sigma^2_{X_j} + b^2_j \sigma^2_Z }
    ,
    \hspace{15pt}
    \kappa_j = \frac{ \kappa_{X_j} \kappa_Z }{ \kappa_{X_j} + \kappa_Z }
    ,
    \hspace{15pt}
    j = 1, ..., n
    .
\end{equation}
The resulting multivariate process $\boldsymbol{Y}(t)$ is then a BB-VG$( \boldsymbol{\mu}_{X}, \boldsymbol{\sigma}_X, \boldsymbol{\kappa}_X, \boldsymbol{b}, \mu_Z, \sigma_Z, \kappa_Z )$ with $4n+3$ parameters, characteristic function 
\begin{equation*}
\begin{split}
    \phi_{\boldsymbol{Y}} (\boldsymbol{u}) =
    & \prod_{j=1}^n \left( 1 - i u_j \mu_{X_j} \kappa_{X_j} + \frac{1}{2} u_j^2 \sigma^2_{X_j} \kappa_{X_j} \right)
    ^{\kappa^{-1}_{X_j}} \cdot
    \\
    & \cdot \left[ 1 - i \mu_Z \kappa_Z \sum_{j=1}^n b_j u_j + \frac{1}{2} \sigma^2_Z \kappa_Z \left(\sum_{j=1}^n b_j u_j \right)^2 \right]
    ^{\kappa^{-1}_Z}
    ,
    \hspace{15pt}
    \boldsymbol{u} \in \mathbb{R}^n
    ,
    \end{split}
\end{equation*}
and correlation coefficient
\begin{equation*}
    \rho_{\boldsymbol{Y}} (i,j)
    =
    \frac{
    b_i b_j (\sigma^2_Z + \mu^2_Z \kappa_Z)
    }{
    \sqrt{ (\sigma^2_i + \mu^2_i \kappa_i)
    (\sigma^2_j + \mu^2_j \kappa_j) }
    }.
\end{equation*}

\subsection{Unconstrained BB-Variance Gamma}\label{BB-VG-uncBB}

Removing the convolution conditions in Eqs.\ \eqref{VG-conv-restrictions-1-copy} and \eqref{VG-conv-restrictions-2-copy}, we can simplify the BB-VG construction.
Let
$
X_j \sim \text{VG} (\mu_{X_j}, \sigma_{X_j}, \kappa_{X_j}),
\hspace{3pt}
Z \sim \text{VG} (\mu_Z, \sigma_Z, \kappa_Z)
$,
with characteristic functions as in Eq.\ \eqref{VG-CF-copy}. Then, without making any further assumption, the characteristic function of $Y_j$ is given by
\begin{equation*}
    \phi_{Y_j} (u) =
    \left( 1 - i u \mu_{X_j} \kappa_{X_j} + \frac{1}{2} u^2 \sigma^2_{X_j} \kappa_{X_j} \right)
    ^{-\kappa^{-1}_{X_j}}
    \hspace{1pt}
    \left( 1 - i u b_j \mu_Z \kappa_Z + \frac{1}{2} u^2 b_j^2 \sigma^2_Z \kappa_Z \right)
    ^{-\kappa^{-1}_Z},
\end{equation*}
with $u \in \mathbb{R}$.
The characteristic function of the resulting multivariate process $\boldsymbol{Y}(t)$ follows from \eqref{BB-CF}, and the pairwise correlation coefficient is
\begin{equation*}
    \rho_{\boldsymbol{Y}} (i,j)
    =
    \frac{
    b_i b_j (\sigma^2_Z + \mu^2_Z \kappa_Z)
    }{
    \sqrt{
    \left[ \sigma^2_{X_i} + \mu^2_{X_i} \kappa_{X_i} + b_i^2 (\sigma^2_Z + \mu^2_Z \kappa_Z ) \right]
    \left[ \sigma^2_{X_j} + \mu^2_{X_j} \kappa_{X_j} + b_j^2 (\sigma^2_Z + \mu^2_Z \kappa_Z ) \right]
    }
    }.
\end{equation*}

\subsection{Constrained BB-Normal Inverse Gaussian}\label{BB-NIG}

\cite{ballotta2016multivariate} originally formulated a multivariate NIG with unbiased subordinator, with parametrization similar to BB-VG.
Here, for sake of comparison with LS construction, we present a new version that keeps the same marginal distributions of the constrained LS-NIG.
Let $Y_j \sim \text{NIG} (\beta_j, \delta_j, \gamma_j), \hspace{3pt} X_j \sim \text{NIG} (\beta_{X_j}, \delta_{X_j}, \gamma_{X_j}), \hspace{3pt} Z \sim \text{NIG} (\beta_Z, \delta_Z, \gamma_Z)$,
with characteristic functions as in Eq.\ \eqref{NIG-CF-copy}. Then, to satisfy Eq.\ \eqref{conv-conditions}, the following constraints must hold,
\begin{equation*}
\begin{aligned}
    \beta_j = b_j^{-1} \beta_Z
    ,
    \hspace{15pt}
    j = 1, ..., n,
    \\
    \gamma_j = b_j^{-1} \gamma_Z
    ,
    \hspace{15pt}
    j = 1, ..., n,
\end{aligned}
\end{equation*}
from which we obtain the relationships
\begin{equation*}
    \beta_j = \beta_{X_j} = b_j^{-1} \beta_Z
    ,
    \hspace{15pt}
    \delta_j = \delta_{X_j} + b_j \delta_Z
    ,
    \hspace{15pt}
    \gamma_j = \gamma_{X_j} = b_j^{-1} \gamma_Z
    ,
    \hspace{15pt}
    j = 1, ..., n
    .
\end{equation*}
The resulting multivariate process $\boldsymbol{Y}(t)$ is then a BB-NIG$( \boldsymbol{\beta}_{X}, \boldsymbol{\delta}_X, \boldsymbol{\gamma}_X, \boldsymbol{b}, \beta_Z, \delta_Z, \gamma_Z )$ with $4n+3$ parameters, characteristic function
\begin{equation*}
\begin{split}
    \phi_{\boldsymbol{Y}} (\boldsymbol{u}) =
    & \prod_{j=1}^n
    \exp \left\{ -\delta_{X_j} \left(\sqrt{\gamma^2_{X_j} - \left(\beta_{X_j} + i u_j \right)^2} - \sqrt{\gamma^2_{X_j} - \beta^2_{X_j}}\right) \right\} \cdot
    \\
    & \cdot \exp \left\{ -\delta_Z \left(\sqrt{\gamma^2_Z - \left(\beta_Z + i \sum_{j=1}^n b_j u_j \right)^2} - \sqrt{\gamma^2_Z - \beta^2_Z} \right) \right\}
    ,
    \hspace{15pt}
    \boldsymbol{u} \in \mathbb{R}^n
    ,
\end{split}
\end{equation*}
and correlation coefficient
\begin{equation*}
    \rho_{\boldsymbol{Y}} (i,j)
    =
    \frac{
    b_i b_j \gamma^2_Z \delta_Z (\gamma^2_Z - \beta^2_Z)^{-3/2}
    }{
    \sqrt{ \gamma^2_i \delta_i (\gamma^2_i - \beta^2_i)^{-3/2}
    \:\:
    \gamma^2_j \delta_j (\gamma^2_j - \beta^2_j)^{-3/2} }
    }.
\end{equation*}

\subsection{Unconstrained BB-Normal Inverse Gaussian}\label{BB-NIG-uncBB}

Also for the NIG case, the unconstrained version is simpler than the constrained one and it is reported here below.
Let
$
X_j \sim \text{NIG} (\beta_{X_j}, \delta_{X_j}, \gamma_{X_j}),
\hspace{3pt}
Z \sim \text{NIG} (\beta_Z, \delta_Z, \gamma_Z)
$,
with characteristic functions as in Eq.\ \eqref{NIG-CF-copy}. Then, without making any further assumption, the characteristic function of $Y_j$ is given by
\begin{equation*}
\begin{split}
    \phi_{Y_j} (u) =
    & \exp \left\{ -\delta_{X_j} \left(\sqrt{\gamma^2_{X_j} - (\beta_{X_j} + i u)^2} - \sqrt{\gamma^2_{X_j} - \beta^2_{X_j}}\right) \right\} \cdot
    \\
    \cdot
    & \exp \left\{ -\delta_Z \left(\sqrt{\gamma^2_Z - (\beta_Z + i u b_j)^2} - \sqrt{\gamma^2_Z - \beta^2_Z}\right) \right\}
    ,
    \hspace{15pt}
    u \in \mathbb{R}
    ,
\end{split}
\end{equation*}
with $u \in \mathbb{R}$.
The characteristic function of the resulting multivariate process $\boldsymbol{Y}(t)$ follows from \eqref{BB-CF}, and the pairwise correlation coefficient is
\begin{equation*}
    \rho_{\boldsymbol{Y}} (i,j)
    =
    \frac{
    b_i b_j \gamma^2_Z \delta_Z (\gamma^2_Z - \beta^2_Z)^{-3/2}
    }{
    \underset{m=i,j}{\prod}
    \sqrt{
    \gamma^2_{X_m} \delta_{X_m} (\gamma^2_{X_m} - \beta^2_{X_m})^{-3/2}
    + b_m^2
    \gamma^2_Z \delta_Z (\gamma^2_Z - \beta^2_Z)^{-3/2}
    }
    }.
\end{equation*}

\newpage
\section{WP3 Discounted Payoff Algorithm}\label{app:WP3 algorithm}

\begin{algorithm}
\caption{WP3 Discounted Payoff Algorithm}\label{alg:WP3-payoff}
\begin{algorithmic}[1]
\State Initialize discounted payoff: $\Pi \gets 0$
\State Initialize counter: $c \gets 0$
\For{$i = 1,...,m$}
    \State $P_j (t_i) \gets S_j (t_i) / S_j (t_0)$, $\forall j = 1,...,n$
    \State $w_i \gets \underset{j \in \{1, ..., n\} }{ \text{argmin}} \{P_j (t_i)\}$
    \If{$P_{w_i} (t_i) \geq b_{d,i}$}
        \State $\Pi \gets \Pi + e^{-r t_i} (c+1) k$
        \State $c \gets 0$
    \Else
        \State $c \gets c + 1$
    \EndIf
    \If{$P_{w_i} (t_i) \geq b_{r,i}$}
        \State $\Pi \gets \Pi + e^{-r t_i} I$
        \State \Return $\Pi$
    \EndIf
\EndFor
\State $P_j (t_m) \gets S_j (t_m) / S_j (t_0)$, $\forall j = 1,...,n$
\State $w_m \gets \underset{j \in \{1, ..., n\} }{ \text{argmin}} \{P_j (t_m)\}$
\If{$P_{w_m} (t_m) \geq b$}
    \State $\Pi \gets \Pi + e^{-r t_m} I$
\Else
    \State $\Pi \gets \Pi + e^{-r t_m} P_{w_m}(t_m) I$
\EndIf
\State \Return $\Pi$
\end{algorithmic}
\end{algorithm}

\end{document}